\PassOptionsToPackage{dvipsnames}{xcolor}
\documentclass[11pt]{article}
\usepackage{jheppub}
\usepackage{bm,bbm}
\usepackage{booktabs,amsmath}
\usepackage{mathtools}
\usepackage{graphics}
\usepackage{braket}
\usepackage{tikz}
\usepackage{ascmac}
\usepackage{enumitem}
\usepackage{comment}
\usepackage{stackrel}
\usepackage{accents}

\usepackage{framed}
\usepackage{colortbl}
\definecolor{shadecolor}{rgb}{0.90,0.90,0.90}
\usetikzlibrary{patterns}
\usetikzlibrary{decorations.markings}
\usetikzlibrary{decorations.pathmorphing}
\tikzset{snake it/.style={decorate, decoration=snake}}

\makeatletter
\newcommand{\mylabel}[2]{#2\def\@currentlabel{#2}\label{#1}}
\makeatother

\newcommand{\RomanNumeralCaps}[1]
    {\MakeUppercase{\romannumeral #1}}

\usetikzlibrary{quotes,angles}

\usetikzlibrary{shadings}

\tikzset{test/.style n args={3}{
    postaction={
    decorate,
    decoration={
    markings,
    mark=between positions 0 and \pgfdecoratedpathlength step 0.5pt with {
    \pgfmathsetmacro\myval{multiply(
        divide(
        \pgfkeysvalueof{/pgf/decoration/mark info/distance from start}, \pgfdecoratedpathlength
        ),
        100
    )};
    \pgfsetfillcolor{#3!\myval!#2};
    \pgfpathcircle{\pgfpointorigin}{#1};
    \pgfusepath{fill};}
}}}}

\definecolor{c1}{rgb}{0.0,0.2,0.9}
\definecolor{c2}{rgb}{0.45,0.0,0.45}
\definecolor{c3}{rgb}{0.9,0.2,0.0}

\usetikzlibrary{arrows}
\usetikzlibrary {arrows.meta}

\tikzset{middlearrow/.style={
        decoration={markings,
            mark= at position 0.57 with {\arrow{#1}} ,
        },
        postaction={decorate}
    }
}

\usepackage{blkarray}

\usepackage{nicematrix}
\NiceMatrixOptions%
{code-for-first-row = \scriptstyle,
code-for-first-col = \scriptstyle }

\usetikzlibrary{decorations.markings}
\def\MarkLt{4pt}
\def\MarkSep{2pt}

\tikzset{
  TwoMarks/.style={
    postaction={decorate,
      decoration={
        markings,
        mark=at position #1 with
          {
              \begin{scope}[xslant=0.2]
              \draw[line width=\MarkSep,white,-] (0pt,-\MarkLt) -- (0pt,\MarkLt) ;
              \draw[-] (-0.5*\MarkSep,-\MarkLt) -- (-0.5*\MarkSep,\MarkLt) ;
              \draw[-] (0.5*\MarkSep,-\MarkLt) -- (0.5*\MarkSep,\MarkLt) ;
              \end{scope}
          }
       }
    }
  },
  TwoMarks/.default={0.5},
  OneMark/.style={
    postaction={decorate,
      decoration={
        markings,
        mark=at position #1 with
          {
              \draw[-] (0,-\MarkLt) -- (0,\MarkLt) ;
          }
       }
    }
  },
  OneMark/.default={0.5}
}

\edef\restoreparindent{\parindent=\the\parindent\relax}
\usepackage{parskip}
\restoreparindent
\usepackage{ascmac}
\usepackage{mathrsfs}
\usepackage{enumitem}
\newlist{steps}{enumerate}{1}
\setlist[steps, 1]{label = Step \arabic*:}

\renewcommand*\arraystretch{1.2}

\def\d{{\rm d}}

\def\CO{{\cal O}}

\def\CW{{\cal W}}

\def\b0{\bm{0}_\perp}

\usepackage{centernot}
\usepackage{mathtools}
\usepackage{stmaryrd}

\makeatletter
\newcommand{\xMapsto}[2][]{\ext@arrow 0599{\Mapstofill@}{#1}{#2}}
\def\Mapstofill@{\arrowfill@{\Mapstochar\Relbar}\Relbar\Rightarrow}
\makeatother

\usepackage{multirow}

\newtheorem{axiom_}{Axiom}
{\begin{axiom_}\begin{shaded}}%
{\end{shaded}\end{axiom_}}

\usepackage{tikz-3dplot}

\DeclareFontFamily{U}{mathx}{\hyphenchar\font45}
\DeclareFontShape{U}{mathx}{m}{n}{
      <5> <6> <7> <8> <9> <10>
      <10.95> <12> <14.4> <17.28> <20.74> <24.88>
      mathx10
      }{}
\DeclareSymbolFont{mathx}{U}{mathx}{m}{n}
\DeclareFontSubstitution{U}{mathx}{m}{n}
\DeclareMathAccent{\widecheck}{0}{mathx}{"71}
\title{Comments on epsilon expansion of the O$(N)$ model with boundary}

\author[a]{Tatsuma Nishioka,}
\author[a,b]{Yoshitaka Okuyama}
\author[a]{and Soichiro Shimamori}

\affiliation[a]{
Department of Physics, Osaka University,\\
Machikaneyama-Cho 1-1, Toyonaka 560-0043, Japan
}

\affiliation[b]{Department of Physics, Faculty of Science,
The University of Tokyo,\\
Bunkyo-Ku, Tokyo 113-0033, Japan}

\preprint{OU-HET-1162}

\abstract{
The O$(N)$ vector model in the presence of a boundary has a non-trivial fixed point in $(4-\epsilon)$ dimensions and exhibits critical behaviors described by boundary conformal field theory.
The spectrum of boundary operators is investigated at the leading order in the $\epsilon$-expansion by diagrammatic and axiomatic approaches.
In the latter, we extend the framework of Rychkov and Tan for the bulk theory to the case with a boundary and calculate the conformal dimensions of boundary composite operators with attention to the analyticity of correlation functions.
In both approaches, we obtain consistent results. 
}

\begin{document}
\maketitle

\section{Introduction}
Quantum field theories (QFTs) with boundaries have been explored since the 1970s with the hope of discovering new phenomena which manifest themselves due to the boundary effects. (See \cite{Diehl:1996kd, binder1983} for older works and \cite{Andrei:2018die} for more recent developments.)
Introducing a boundary breaks a part of the symmetries in QFTs, such as translation and rotation.
When the bulk theory is tuned to a critical point described by a conformal field theory (CFT), the conformal symmetry is broken to a subgroup by the boundary and there appear to be no critical behaviors. Nevertheless, boundary critical phenomena are characterized by boundary CFTs (BCFTs) if the shape of the boundary is planar or spherical and a proper boundary condition is imposed.
Depending on the choice of boundary conditions, there are several BCFTs associated with the same bulk CFT.
To analyze boundary critical phenomena in $(1+1)$ dimensions, Cardy developed novel techniques in two-dimensional BCFTs (BCFT$_2$) \cite{Cardy:1984bb} and classified admissible boundary conditions \cite{Cardy:1989ir}.
Furthermore, McAvity and Osborn generalized Cardy's discussion on BCFT$_2$ to higher-dimensional BCFTs (BCFT$_{d\, \geq 3}$) \cite{McAvity:1993ue,McAvity:1995zd}, which provides us indispensable techniques to investigate the boundary effects in CFT as we will see below.

In BCFT$_{d\geq 3}$, the boundary breaks the full conformal symmetry SO$(1,d+1)$ down to its subgroup SO$(1,d)$.\footnote{Throughout this paper, we treat Euclidean spacetime.}
Further, we distinguish operators in the bulk from those on the boundary.
The former and latter are called bulk and boundary local operators respectively. 
Bulk local operators are characterized by the representation of SO$(1,d+1)$ in the same way as in CFTs without boundaries and defects.
On the other hand, boundary local operators represent dynamical degrees of freedom localized on the boundary and constitute conformal families of the subgroup SO$(1,d)$.
These two types of conformal families are not independent, and one can expand a bulk operator in terms of boundary ones, which is known as boundary operator expansion (BOE). 
For a bulk local operator $\CO_{\Delta}$ of conformal dimension $\Delta$, the BOE reads \cite{McAvity:1995zd}:
\begin{align}\label{eq: BOE general}
\CO_{\Delta}(x) = \sum_{\widehat{\CO}}\, C_{\Delta,\widehat{\Delta}}(x_\perp, \widehat{\partial}\, )\, \widehat{\CO}_{\widehat{\Delta}}(\hat{x})\, ,\qquad x=(\hat{x}^{a}, x_\perp)\, ,\qquad a=1,\cdots, d-1\ . 
\end{align}
We use coordinates $\hat{x}$/$x_\perp$ for the parallel/transverse directions to the boundary respectively. The sum is taken over all possible boundary primary operators $\widehat{\CO}_{\widehat{\Delta}}$
and depends on the choices of the boundary conditions. The differential operator $C_{\Delta,\widehat{\Delta}}(x_\perp, \widehat{\partial})$ is completely determined by the conformal symmetry modulo model-dependent coefficients.

In this paper, we will concentrate at the Wilson-Fisher fixed point of the $\text{O}(N)$ vector model in $d=(4-\epsilon)$-dimensional half-spacetime $\mathbb{R}^{d}_{+} \equiv \mathbb{R}^{d-1}\times \mathbb{R}_{\geq 0}$ \cite{McAvity:1993ue,McAvity:1995zd}:
\begin{align}\label{eq: action}
    I = \int_{\mathbb{R}^{d}_{+}}\,\d^d x \,\left[\frac{1}{2\,(d-2)\,\Omega_{d-1}}\, |\partial \Phi_{1} |^2\, +\,   \frac{\lambda\, \mu^{\epsilon}}{4!}\, |\Phi_1|^4 \right] \, ,\qquad |\Phi_1|^4 \equiv (\Phi_{1}^{\alpha}\Phi_{1}^{\alpha})^2\ . 
\end{align}
Here $\Phi_{1}^{\alpha}$ is an O$(N)$ vector field subject to either Neumann or Dirichlet boundary condition, and $\Omega_{d-1}$ is the volume of a $(d-1)$-dimensional sphere: $\Omega_{d-1}=2\pi^{d/2}/\Gamma(d/2)$. The beta function $\beta_{\lambda}$ for the bulk coupling $\lambda$ is (see e.g., \cite{Peskin:1995ev})
 \begin{align}\label{eq:beta function}
     \beta_{\lambda}=\frac{\d \lambda}{\d \log \mu} =-\epsilon\, \lambda + \frac{N+8}{3}\,\pi^2\lambda^2+O(\lambda^3) \, ,
 \end{align}
 where $\mu$ is a momentum scale. This O$(N)$ model exhibits critical behaviors at the infrared fixed point with
\begin{align}\label{eq: Wilson-Fisher FP}
    \lambda_\ast\equiv \frac{1}{\pi^2}\frac{3}{N+8}\, \epsilon\, +\, O(\epsilon^2)\, .
\end{align}

The conformal dimensions of bulk operators at the Wilson-Fisher fixed point can be derived in perturbation theory \cite{Wilson:1971dc,Wilson:1973jj,Collins:1984xc,Peskin:1995ev, Kleinert:2001ax,kardar_2007}.
Rychkov and Tan reproduced the leading anomalous dimensions by the axiomatic method \cite{Rychkov:2015naa}, which compare the free O$(N)$ theory in four dimensions and the Wilson-Fisher fixed point in $(4-\epsilon)$ dimensions without resorting to diagrammatic calculations.
The validity of their method was examined in the $\phi^4$-theory, and successive studies revealed that their framework can be applied to various models including not only bosonic but also fermionic fields \cite{Basu:2015gpa, Ghosh:2015opa, Raju:2015fza, Sen:2015doa, Nii:2016lpa, Roumpedakis:2016qcg, Giombi:2016hkj, Gopakumar:2016cpb, Gopakumar:2016wkt, Hasegawa:2016piv, Gliozzi:2016ysv}. 
Furthermore, this method was adapted to defect CFT with a monodromy defect \cite{Billo:2013jda, Gaiotto:2013nva} in $(4-\epsilon)$ dimensions \cite{Yamaguchi:2016pbj, Soderberg:2017oaa}, which precisely reproduced the anomalous dimensions of the defect operators. 
More recently, the Rychkov-Tan approach has been combined with the conformal bootstrap \cite{Rattazzi:2008pe,Poland:2018epd} to derive new constraints on the dynamics of fermionic boundary and defect CFTs, such as Yukawa and Gross-Neveu model, etc. \cite{Herzog:2022jlx,Giombi:2022vnz}.\footnote{See e.g., \cite{Liendo:2012hy,Bissi:2018mcq,Dey:2020jlc,Padayasi:2021sik,Gliozzi:2015qsa,Gimenez-Grau:2022ebb,Gimenez-Grau:2022czc,Bianchi:2022sbz} for the applications of conformal bootstrap to the O$(N)$ vector models with boundaries or defects.}

A few comments on our notation of operators are in order. (See table \ref{Table: notations of operators}.)
We focus on the Neumann boundary condition in the main text and leave the Dirichlet case to appendix \ref{app:Dirichlet boundary condition}.
\begin{table}[t]
\renewcommand{\arraystretch}{1.5}
\centering
\begin{tabular}{>{\centering}m{3cm}>{\centering}m{3cm}>{\centering\arraybackslash}m{4cm}} \toprule
     & Bulk local op. & Boundary local op. \\  \midrule
    Free & $\Phi_{2p}$ ,  $\Phi_{2p+1}^{\alpha}$ & $\widehat{\Phi}_{2p}$ ,  $\widehat{\Phi}_{2p+1}^{\alpha}$\\
    Wilson-Fisher& $W_{2p}$ , $W_{2p+1}^{\alpha}$ & $\widehat{W}_{2p}$ ,  $\widehat{W}_{2p+1}^{\alpha}$\\ \bottomrule
\end{tabular}
\caption{List of symbols for bulk and boundary local operators. The lower indices of the operators denote the canonical dimensions in $d=4$.}
\label{Table: notations of operators}
\end{table}
The subscripts of operators indicate the canonical dimensions in $d=4$. 
We use different symbols to distinguish bulk/boundary local operators in the free theory from those at the Wilson-Fisher fixed point. 
Two sets of operators $\{\Phi_{2p}\equiv|\Phi_{1}|^{2p},\Phi_{2p+1}^{\alpha}\equiv \Phi_{1}^{\alpha}\,|\Phi_{1}|^{2p}\}$ and $\{\widehat{\Phi}_{2p}, \widehat{\Phi}_{2p+1}^{\alpha}\}$ stand for the bulk and boundary operators in the free theory, respectively.
Under the Neumann boundary condition, $\{\widehat{\Phi}_{2p},\widehat{\Phi}_{2p+1}^{\alpha}\}$ are defined as follows:
\begin{align}
    \widehat{\Phi}_{2p}(\hat{x})=\lim_{x_{\perp}\rightarrow 0}|\Phi_{1}|^{2p}(x)\, , \qquad \widehat{\Phi}_{2p+1}^{\alpha}(\hat{x})=\lim_{x_{\perp}\rightarrow 0}\Phi_{1}^{\alpha}\,|\Phi_{1}|^{2p}(x)\, .
\end{align}
On the other hand, $\{W^{\alpha}_{2p},W^{\alpha}_{2p+1}\}$ and $\{\widehat{W}_{2p},\widehat{W}_{2p+1}^{\alpha}\}$ are the renormalized bulk and boundary operators at the Wilson-Fisher fixed point, which in the free limit tend to
\begin{align}
    \left\{ W_{2p}\ ,\ W_{2p+1}^{\, \alpha}\ ,\ \widehat{W}_{2p}\ ,\ \widehat{W}_{2p+1}^{\, \alpha} \right\} \quad \xrightarrow[]{\text{free}} \quad \left\{ \Phi_{2p}\ ,\  \Phi_{2p+1}^{\, \alpha}\ ,\  \widehat{\Phi}_{2p}\ ,\  \widehat{\Phi}_{2p+1}^{\, \alpha} \right\}\ . 
\end{align}
By using the Rychkov-Tan method, the anomalous dimension of the lowest-lying boundary operator $\widehat{W}_{1}^{\, \alpha}$ has already been examined in \cite{Dey:2020jlc,Giombi:2020rmc}. To our best knowledge, however, the anomalous dimensions of boundary composite operators have not yet been derived by the axiomatic approach.
\paragraph{Summary and discussion.} In this paper, we reproduce the leading anomalous dimensions of boundary composite operators by the Rychkov-Tan method,\footnote{It remains open how to extract the higher-order anomalous dimensions by the Rychkov-Tan method, even in the absence of a boundary.}
and verify that the results are the same as the diagrammatic ones. 
We focus on Neumann and Dirichlet boundary conditions which correspond to special and ordinary transitions respectively \cite{Diehl:1996kd}.\footnote{There is one more critical phase called the extraordinary transition \cite{Lubensky19752, Bray1977} which can be described by BCFT.
In this phase, unlike the special and ordinary ones, there is no corresponding phase in four dimensions ($\epsilon=0$) (see e.g., appendix B.4 in \cite{Liendo:2012hy}).
Thus, we cannot apply our axiomatic method to examine the extraordinary phase.}
We find it necessary to take into account the analyticity of correlation functions to calculate the anomalous dimensions of boundary composite operators.
We believe that our method presented in this paper is also applicable to other boundary composite operators including derivatives. 
Furthermore, we expect that our argument is not limited to the O$(N)$ vector model with a boundary but can also be applied to similar models with a line defect \cite{Allais:2014fqa,Cuomo:2021kfm} as demonstrated in the accompanying paper \cite{Nishioka:2022qmj}, and presumably to other models with boundaries or defects.
We hope to investigate this direction in the future.

This paper is organized as follows. In section \ref{sec:review of BCFT}, we review the basics of BCFT. In section \ref{sec:Correlators in free theory Neumann}, we provide various aspects of the free O$(N)$ model with the Neumann boundary condition. 
In section \ref{Diagrammatic calculation, Neu}, we perform diagrammatic calculations to derive the anomalous dimensions of the boundary composite operators $\widehat{W}_{2p}$ and $\widehat{W}_{2p+1}^{\, \alpha}$.
In section \ref{sec:Rychkov-Tan Axioms}, we use the Rychkov-Tan method to reproduce the diagrammatic results obtained in section \ref{Diagrammatic calculation, Neu}.
The resulting conformal dimensions are summarized in table \ref{TableEps2}.
In appendix \ref{app:Dirichlet boundary condition}, we derive the anomalous dimensions of composite operators subject to the Dirichlet boundary condition.
\begin{table}[t]
\renewcommand{\arraystretch}{2.2}
\centering
\begin{tabular}{>{\centering}m{2.7cm}>{\centering}m{2.5cm}>{\centering}m{6cm}>{\centering\arraybackslash}m{2.5cm}}
\toprule
    Boundary condition & Boundary operators & 
    Conformal dimension & Free limit 
    \\
  \midrule 
    &$\widehat{W}_{2p}$  & $\displaystyle 2p +\frac{6p\, (2p-3)}{N+8}\,\epsilon$ &$\widehat{\Phi}_{2p}$  \\
    \multirow{-2}{*}{Neumann}&$\widehat{W}_{2p+1}^{\,\alpha}$  & $\displaystyle  2p+1-\frac{N+6p\,(1-2p)+5}{N+8}\,\epsilon$ &$\widehat{\Phi}_{2p+1}^{\,\alpha}$  \\
    \hline
    &$\widehat{\CW}_{4p}$  & $\displaystyle 4p-\frac{p\, (N-6p+14)}{N+8}\,\epsilon$ &$\widehat{\Psi}_{4p}$  \\
    \multirow{-2}{*}{Dirichlet}&$\widehat{\CW}_{4p+2}^{\,\alpha}$  & $\displaystyle  4p+2-\frac{N-6p^2 +p\,(N+8)+5}{N+8}\,\epsilon$ &$\widehat{\Psi}_{4p+2}^{\,\alpha}$  \\
    \bottomrule
\end{tabular}
\caption{The conformal dimensions of boundary local operators subject to the Neumann and Dirichlet boundary conditions. See appendix \ref{app:Dirichlet boundary condition} for the notations for the Dirichlet case. 
}
\label{TableEps2}
\end{table}

\section{Review of boundary conformal field theory}\label{sec:review of BCFT}
Before going to the specific model, we record the consequences of conformal symmetry in the presence of a boundary \cite{Billo:2016cpy,Gadde:2016fbj,Cardy:1984bb,McAvity:1995zd}. 

\paragraph{Notations.}
To set the stage, we first introduce our notations. 
Throughout this paper, we consider quantum field theories on the $d$-dimensional Euclidean flat manifold with a boundary: $\mathbb{R}_{+}^{d}=\mathbb{R}^{d-1}\times \mathbb{R}_{\geq 0}$.
Throughout this paper, symbols with hat (\,$\widehat{\,\, }$\,) such as $\widehat \CO$ and $\hat y$ notate the boundary quantities while those without hat are for the bulk ones.

We use $x^\mu$ ($\mu=1,\cdots,d$) for the bulk coordinates, which can be split into the parallel ($\hat x^a$) and transverse ($x_\perp$) parts to the boundary such that $x^\mu\equiv(\hat{x}^{a}, x_\perp)$ with $a=1,\cdots, d-1$ and $x_\perp\geq0$.
Then, $\hat y^\mu\equiv(\hat{y}^a,0)$ stands for the boundary coordinates.
The distances between bulk-bulk, bulk-boundary and boundary-boundary points are written as
\begin{align}
    \begin{aligned}
        |x_{1}-x_{2}|^2 &\equiv (x_{1}-x_{2})^{\mu}\, (x_{1}-x_{2})_{\mu}\ , \\
        |x-\hat{y}|^2 &\equiv (\hat{x}-\hat{y})^{a}\, (\hat{x}-\hat{y})_{a}+x_{\perp}^2\ , \\
        |\hat{y}_{1}-\hat{y}_{2}|^2 &\equiv (\hat{y}_{1}-\hat{y}_{2})^{a}\, (\hat{y}_{1}-\hat{y}_{2})_{a}\ .
    \end{aligned}
\end{align}
A bulk (boundary) scalar primary operator with conformal dimension $\Delta$ ($\widehat{\Delta}$) will be denoted by $\CO_\Delta$ ($\widehat{\CO}_{\widehat{\Delta}}$).
The following shorthanded notations are used frequently in this paper:
 \begin{align}
\hat{y}_{ij}\equiv\hat{y}_i-\hat{y}_j\ ,\qquad \widehat{\Delta}^{\pm}_{ij}\equiv\widehat{\Delta}_i\pm \widehat{\Delta}_j\ .
 \end{align}
 
\paragraph{Correlation functions and boundary operator expansions.}

The bulk-boundary and boundary two-point functions and the boundary three-point function can be fixed by boundary conformal symmetry:
\begin{align}
    \langle\,\CO_{\Delta}(x)\,\widehat{\CO}_{\widehat{\Delta}}(\hat{y})\,\rangle
        &=
            \frac{b(\CO,\widehat{\CO})}{|x-\hat{y}|^{2\widehat{\Delta}}\,|x_\perp|^{\Delta-\widehat{\Delta}}}\ ,\label{eq:btb 2pt gen}\\
            \langle\,\widehat{\CO}_{\widehat{\Delta}}(\hat{y}_1)\,\widehat{\CO}_{\widehat{\Delta}}(\hat{y}_2)\,\rangle
            &=
                \frac{c(\widehat{\CO},\widehat{\CO})}{|\hat{y}_{12}|^{2\widehat{\Delta}}}\ ,\label{eq:bb 2pt gen} \\
            \langle\,\widehat{\CO}_{\widehat{\Delta}_1}(\hat{y}_1)\,\widehat{\CO}_{\widehat{\Delta}_2}(\hat{y}_2) \,\widehat{\CO}_{\widehat{\Delta}_3}(\hat{y}_3) \,\rangle
        &=
            \frac{c(\widehat{\CO}_1,\widehat{\CO}_2,\widehat{\CO}_2)}{|\hat{y}_{12}|^{\widehat{\Delta}^+_{12}-\widehat{\Delta}_3}\,|\hat y_{23}|^{\widehat{\Delta}^+_{23}-\widehat{\Delta}_1}\,|\hat y_{13}|^{\widehat{\Delta}^+_{13}-\widehat{\Delta}_2}}\label{eq:bbb 3pt gen}\ .
\end{align}
The BOE of a bulk scalar operator $\CO_{\Delta}$ is
\begin{align}\label{eq:btb OPE schematic}
    \CO_{\Delta}(x)\supset\sum_{\widehat{\CO}} \,\frac{b(\CO,\widehat{\CO})/c(\widehat{\CO},\widehat{\CO})}{|x_\perp|^{\Delta-\widehat{\Delta}}}\,\widehat{\CO}_{\widehat{\Delta}}(\hat{x})\ .
\end{align}
Here, $\supset$ stands for the BOE, and we do not explicitly write down descendant terms.

The bulk-boundary-boundary three-point functions are not completely fixed by conformal symmetry, and they admit the following conformal block expansion \cite{Lauria:2020emq,Nishioka:2022qmj}:
\begin{align}\label{eq:conformal block expansion main}
              \langle\, \CO_{\Delta}(x)\,\widehat{\CO}_{\widehat{\Delta}_1}(0)\,\widehat{\CO}_{\widehat{\Delta}_2}(\infty) \,\rangle=
                    \frac{1}{|x_\perp|^{\Delta} \,|x|^{\widehat{\Delta}^-_{12}}}\, \sum_{\widehat{\CO}}\, \frac{b(\CO,\widehat{\CO})\,c(\widehat{\CO},\widehat{\CO}_1,\widehat{\CO}_2)}{c(\widehat{\CO},\widehat{\CO})}\,G^{\widehat{\Delta}^-_{12}}_{\widehat{\Delta}}\left(\frac{|x_\perp|^2}{|x|^2}\right)\ .
\end{align}
Here, we set $y_1=0,\ y_2 =\infty$ by using the boundary conformal symmetry.\footnote{Note that $\widehat{\CO}_{\widehat{\Delta}}(\infty)=\lim_{|y|\to\infty}\,|\hat{y}|^{2\widehat{\Delta}}\,\widehat{\CO}_{\widehat{\Delta}}(\hat{y})$.}
The conformal block $G^{\widehat{\Delta}^-_{12}}_{\widehat{\Delta}}(\upsilon)$ can be written as \cite{Nishioka:2022qmj}:
\begin{align}\label{eq:conformal block bdy}
    G^{\widehat{\Delta}^-_{12}}_{\widehat{\Delta}}(\upsilon)
        =
            \upsilon^{\widehat{\Delta}/2}\,{}_2F_1\left(\frac{\widehat{\Delta}+\widehat{\Delta}^-_{12}}{2},\frac{\widehat{\Delta}-\widehat{\Delta}^-_{12}}{2};\widehat{\Delta}-\frac{d-3}{2};\upsilon\right)\ .
\end{align}

\section{The free $\textrm{O}(N)$ model with Neumann boundary condition}\label{sec:Correlators in free theory Neumann}
We investigate the conformal structures of the free O$(N)$ vector model with the Neumann boundary condition in $d$ dimensions with special attention to correlation functions and BOEs, which will play a central role in the rest of the paper.

\subsection{Correlation functions in arbitrary dimensions}\label{eq:Correlation functions in arbitrary dimensions Neu}
We start with the bulk two-point function of the fundamental scalar field satisfying the following differential equation:
\begin{align}\label{eq:4d free scalar propagator EoM}
    \Box_{x_1}\,\langle\,\Phi_{1}^\alpha(x_1)\,\Phi_{1}^\beta(x_2)\,\rangle
        =
            \frac{4\pi^{d/2}}{\Gamma(d/2)}\,\delta^{\alpha\beta}\,\delta^d(x_1-x_2)\ .
\end{align}
Under the Neumann boundary condition
\begin{align}
    \lim_{x_\perp\to0}\,\frac{\partial}{\partial x_\perp}\,\Phi_1^\alpha(x)=0\ ,
\end{align}
the solution to the differential equation \eqref{eq:4d free scalar propagator EoM} is given by
\begin{align}\label{eq:4d free scalar propagator Neumann}
    \langle\,\Phi_{1}^\alpha(x_1)\,\Phi_{1}^\beta(x_2)\,\rangle
        =
            \delta^{\alpha\beta}\,\left[\frac{1}{|x_1-x_2|^{d-2}}+ \frac{1}{|x_1-\bar{x}_2|^{d-2}}\right]\ ,\qquad 
    \bar{x}^\mu
        =
            (\hat{x}^a,-x_\perp)\ .
\end{align}
From this expression, we obtain the bulk one-point functions of the composite operators:
\begin{align}\label{eq:N bulk 1pt}
  \langle\,\Phi_{1}^\alpha\Phi_{1}^\beta(x)\,\rangle
    =
        \frac{\delta^{\alpha\beta}}{2^{d-2}\,|x_\perp|^{d-2}}\ ,\qquad   \langle\,|\Phi_1|^{2}(x)\,\rangle
    =
        \frac{N}{2^{d-2}\,|x_\perp|^{d-2}}\ .
\end{align}
In addition, letting $x_\perp\to0$ in the bulk two-point function \eqref{eq:4d free scalar propagator Neumann}, we find the following bulk-boundary and boundary-boundary functions:
\begin{align}\label{eq:2pt Phi Neu}
    \langle\,\Phi_1^\alpha(x)\,\widehat{\Phi}_{1}^{\,\beta}(\hat{y})\,\rangle
        =
            \frac{2\, \delta^{\alpha\beta}}{|x-\hat{y}|^{d-2}} \ , \qquad
       \langle\,\widehat{\Phi}_{1}^{\,\alpha}(\hat{y}_1)\,\widehat{\Phi}_{1}^{\,\beta}(\hat{y}_2)\,\rangle
        =
            \frac{2\, \delta^{\alpha\beta}}{|\hat{y}_{12}|^{d-2}} \ .
\end{align}
Combined with the results obtained so far, we can calculate any correlators in the free theory by using Wick's theorem.
For instance, we have
\begin{align}\label{eq:Neu btb 2pt Phi3}
\begin{aligned}
      \langle\,\Phi_3^\alpha(x)\,\widehat{\Phi}_{1}^{\,\beta}(\hat{y})\,\rangle
        &=
            \frac{(N/2+1)\,\delta^{\alpha\beta}}{2^{d-4}\,|x-\hat{y}|^{d-2}\,|x_\perp|^{d-2}}\ ,\\
            \langle\,\Phi_3^\alpha(x)\,\widehat{\Phi}_{3}^{\,\beta}(\hat{y})\,\rangle
                &=
                \frac{  32\,(N/2+1)\,\delta^{\alpha\beta}}{|x-\hat{y}|^{3(d-2)}}\ ,\\
   \langle\,\widehat{\Phi}_{3}^\alpha(\hat{y}_1)\,\widehat{\Phi}_{3}^{\,\beta}(\hat{y}_2)\,\rangle
                        &=
                        \frac{  32\,(N/2+1)\,\delta^{\alpha\beta}}{|\hat{y}_{12}|^{3(d-2)}}\ .
\end{aligned}
\end{align}
Furthermore, the two-point functions of the boundary composite operators are given by
\begin{align}
                       \langle\, \widehat{\Phi}_{2p}(\hat{y}_1)\,\widehat{\Phi}_{2p}(\hat{y}_2) \,\rangle
                        &=
                        \frac{N\, g_{p-1}}{|\hat{y}_{12}|^{2p(d-2)}}\ ,\qquad
                                                   \langle\, \widehat{\Phi}_{2p+1}^{\,\alpha}(\hat{y}_1)\,\widehat{\Phi}_{2p+1}^{\,\beta}(\hat{y}_2) \, \rangle
                        =
                        \frac{f_p\, \delta^{\alpha\beta}}{|\hat{y}_{12}|^{(2p+1)(d-2)}}\ ,
\end{align}
where we introduced two combinatorial factors $f_p$ and $g_p$\footnote{Here we use the Pochhammer symbol: $(u)_{n}\equiv \Gamma(u+n)/\Gamma(u)$.}:
\begin{align}\label{eq:Neumann 3pf coeff f g}
    f_p=2^{4p+1}\,p!\, (N/2+1)_p\ ,\qquad g_p=2^{4p+3}\,(p+1)!\, (N/2+1)_p\ .
\end{align}

In what follows, we enumerate several three-point functions of our interest without derivation.

\paragraph{Boundary three-point functions.}
    \begin{align}
     \langle\,\widehat{\Phi}_1^\alpha(\hat{x})\, \widehat{\Phi}_{2p}(\hat{y}_1)\,\widehat{\Phi}_{2p+1}^{\,\beta}(\hat{y}_2)\,\rangle&=\frac{f_p\, \delta^{\alpha\beta}}{|\hat{x}-\hat{y}_2|^{d-2}\,|\hat{y}_{12}|^{2p(d-2)}}\ ,\label{eq:3pt ddd Neumann 1}\\
     \langle\,\widehat{\Phi}_1^\alpha(\hat{x})\, \widehat{\Phi}_{2p+1}^{\beta}(\hat{y}_1)\,\widehat{\Phi}_{2p+2}(\hat{y}_2)\,\rangle
    &=   \frac{g_p\, \delta^{\alpha\beta}}{|\hat{x}-\hat{y}_2|^{d-2}\,|\hat{y}_{12}|^{(2p+1)(d-2)}}\ ,\label{eq:3pt ddd Neumann 2}\\
       \langle\,\widehat{\Phi}_3^\alpha(\hat{x})\, \widehat{\Phi}_{2p}(\hat{y}_1)\,\widehat{\Phi}_{2p+1}^{\,\beta}(\hat{y}_2)\,\rangle&=\frac{ 12p\,f_p\, \delta^{\alpha\beta}}{|\hat{x}-\hat{y}_1|^{d-2}\,|\hat{x}-\hat{y}_2|^{2(d-2)}\,|\hat{y}_{12}|^{(2p-1)(d-2)}} \label{eq:3pt ddd Neumann 3} \ ,\\
     \langle\,\widehat{\Phi}_3^\alpha(\hat{x})\, \widehat{\Phi}_{2p+1}^{\,\beta}(\hat{y}_1)\,\widehat{\Phi}_{2p+2}(\hat{y}_2)\,\rangle&=\frac{2\,(N+6p+2)\,g_p\, \delta^{\alpha\beta}}{|\hat{x}-\hat{y}_1|^{d-2}\,|x-\hat{y}_2|^{2(d-2)}\,|\hat{y}_{12}|^{2p(d-2)}}\ ,\label{eq:3pt ddd Neumann 4}
\end{align}

\paragraph{Bulk-boundary-boundary three-point functions.}
\begin{align}
    \langle\,\Phi_1^\alpha(x)\, \widehat{\Phi}_{2p}(\hat{y}_1)\,\widehat{\Phi}_{2p+1}^{\,\beta}(\hat{y}_2)\,\rangle
        &= 
            \frac{f_p\, \delta^{\alpha\beta}}{|x-\hat{y}_2|^{d-2}\,|\hat{y}_{12}|^{2p(d-2)}}\ ,\label{eq:3pt bdd Neumann 1}\\
    \langle\,\Phi_1^\alpha(x)\, \widehat{\Phi}_{2p+1}^{\beta}(\hat{y}_1)\,\widehat{\Phi}_{2p+2}(\hat{y}_2)\,\rangle
        &=  
            \frac{g_p\, \delta^{\alpha\beta}}{|x-\hat{y}_2|^{d-2}\,|\hat{y}_{12}|^{(2p+1)(d-2)}}\ ,\label{eq:3pt bdd Neumann 2}
    \end{align}
     \begin{align}
    &\langle\,\Phi_3^\alpha(x)\, \widehat{\Phi}_{2p}(\hat{y}_1)\,\widehat{\Phi}_{2p+1}^{\,\beta}(\hat{y}_2)\,\rangle
         \nonumber\\ 
         & \qquad 
         = 
                \frac{N+2}{2^{d-2}}\, \frac{f_p\, \delta^{\alpha\beta}}{|x-\hat{y}_2|^{d-2}\,|\hat{y}_{12}|^{2p(d-2)}\,|x_\perp|^{d-2}}
                 +
                 \frac{12\,p\, f_p\,\delta^{\alpha\beta}}{|x-\hat{y}_1|^{d-2}\,|x-\hat{y}_2|^{2(d-2)}\,|\hat{y}_{12}|^{(2p-1)(d-2)}}
                   \ , \label{eq:3pt bdd Neumann 3}\\
    &\langle\,\Phi_3^\alpha(x)\, \widehat{\Phi}_{2p+1}^{\,\beta}(\hat{y}_1)\,\widehat{\Phi}_{2p+2}(\hat{y}_2)\,\rangle
         \nonumber \\
        & \qquad 
        =\frac{N+2}{2^{d-2}}\,
            \frac{g_p\, \delta^{\alpha\beta}}{|x-\hat{y}_2|^{d-2}\,|\hat{y}_{12}|^{(2p+1)(d-2)}\,|x_\perp|^{d-2}}
             +
          \frac{2\,(N+6p+2)\,g_p\, \delta^{\alpha\beta}}{|x-\hat{y}_1|^{d-2}\,|x-\hat{y}_2|^{2(d-2)}\,|\hat{y}_{12}|^{2p(d-2)                               }}\ .
                \label{eq:3pt bdd Neumann 4}
\end{align}

\subsection{Boundary operator expansions in four dimensions}
We set $d=4$ and examine the boundary operator expansions in the four-dimensional free O$(N)$ model with the Neumann boundary condition. In particular, we focus on the two operators $\Phi_1^\alpha$ and $\Phi_3^\alpha$ which are relevant in section \ref{sec:Rychkov-Tan Axioms}.

\subsubsection{Boundary operator expansion of $\Phi_1^\alpha$}\label{sec:Boundary operator expansion of Phi1}
From the bulk-boundary two-point function \eqref{eq:2pt Phi Neu}, we can deduce that $\Phi_1^\alpha$ has the following BOE:
\begin{align}
    \Phi_1^\alpha(x)        \supset\widehat{\Phi}_{1}^{\,\alpha}(\hat{x})\ .
\end{align}
To see if the other operators contribute to the BOE, we rewrite the bulk-boundary-boundary three-point correlators \eqref{eq:3pt bdd Neumann 1} and \eqref{eq:3pt bdd Neumann 2} in terms of conformal blocks \eqref{eq:conformal block expansion main} using the relation $G^{-1}_{1}(\upsilon)=\upsilon^{1/2}$ as follows:
\begin{align}
     \langle\,\Phi_1^\alpha(x)\, \widehat{\Phi}_{2p}(0)\,\widehat{\Phi}_{2p+1}^{\,\beta}(\infty)\,\rangle
        &=
            f_p\, \delta^{\alpha\beta}\,\frac{|x|}{|x_\perp|}\, G^{-1}_{1}\left(\frac{|x_\perp|^2}{|x|^2}\right)\ ,\label{eq:3pt bdd Neumann 1 exp}\\
      \langle\,\Phi_1^\alpha(x)\, \widehat{\Phi}_{2p+1}^{\beta}(0)\,\widehat{\Phi}_{2p+2}(\infty)\,\rangle
        &=
            g_p\,\delta^{\alpha\beta}\,\frac{|x|}{|x_\perp|}\, G^{-1}_{1}\left(\frac{|x_\perp|^2}{|x|^2}\right)\ .\label{eq:3pt bdd Neumann 2 exp}
\end{align}
Compared with the three-point function \eqref{eq:conformal block expansion main} with a general structure of the BOE \eqref{eq:btb OPE schematic}, there are no contributions to the BOE of $\Phi_1^\alpha$ other than a boundary operator of dimension $\widehat\Delta = 1$, which may be identified with $\widehat{\Phi}_{1}^{\,\alpha}$ in the present case.

We can confirm the above statement in a different manner as follows.
Suppose a boundary scalar primary operator $\widehat{\CO}_{\widehat{\Delta}}(\hat{x})$ appears in the BOE of $\Phi_1^\alpha(x)$:
\begin{align}\label{eq:boundary free scalar OPE}
    \Phi_1^\alpha(x)\supset \frac{A}{|x_\perp|^{1-\widehat{\Delta}}}\, \widehat{\CO}_{\widehat{\Delta}}(\hat{x})\ ,
\end{align}
where $A$ is a non-zero constant.
In the free theory, $\Phi_1^\alpha(x)$ satisfies the Klein-Gordon equation $\Box_x\,\Phi_1^\alpha(x)=0$ which holds as an operator identity.
Applying the Laplacian $\Box_x$ to the both sides of \eqref{eq:boundary free scalar OPE} we find
\begin{align}
   0= \Box_x\,\Phi_1^\alpha(x)\supset (\widehat{\Delta}-1)(\widehat{\Delta}-2)\,\frac{A}{|x_\perp|^{3-\widehat{\Delta}}}\, \widehat{\CO}_{\widehat{\Delta}}(\hat{x})\ .
\end{align}
For this operator identity to hold for a non-zero constant $A$, the conformal dimensions of $\widehat{\CO}_{\widehat{\Delta}}$ must be either one or two.
The former corresponds to the Neumann boundary condition we are considering here, and the latter to the Dirichlet boundary condition.

We thus conclude that with the Neumann boundary condition only $\widehat{\Phi}_1^\alpha$-channel appears in the BOE of $\Phi_1^\alpha$:
\begin{align}
        \Phi_1^\alpha(x)
        &=\widehat{\Phi}_{1}^{\,\alpha}(\hat{x})+(\text{descendants})\ .\label{eq:OPE of phi1 O(N) Neumann}
\end{align}

\subsubsection{Boundary operator expansion of $\Phi_3^\alpha$}
The structure of the BOE of $\Phi_3^\alpha$ is much richer than that of $\Phi_1^\alpha$. From the two-point functions \eqref{eq:Neu btb 2pt Phi3}, we find $\widehat{\Phi}_{1}^{\,\alpha}$ and $\widehat{\Phi}_3^\alpha$ contribute to the BOE:
\begin{align}
   \Phi_3^\alpha(x)
        \supset \frac{N+2}{4\,|x_\perp|^2}\,\widehat{\Phi}_{1}^{\,\alpha}(\hat{x})+\widehat{\Phi}_3^\alpha(\hat{x}) \label{eq:OPE of phi3 O(N) Neumann}\ .
\end{align}
There are also other (infinitely many) boundary operators that appear in the BOE of $\Phi_3^\alpha$. To see this, let us perform the conformal block expansion of the bulk-boundary-boundary three-point functions involving $\Phi_3^\alpha$.
Using the boundary conformal symmetry, we can place the two boundary operators at $\hat{x} = 0$ and $\hat{x} = \infty$.
Then, \eqref{eq:3pt bdd Neumann 3} yields\footnote{We used $G^{-1}_{1}(\upsilon)=\upsilon^{1/2}$ and the following identity for hypergeometric functions \cite[equation (9.1.32)]{luke1969special}:
\begin{align*}
1=\sum_{n=0}^{\infty}\,\frac{(-1)^n\,(\alpha)_n(\beta)_n}{(n+\lambda)_n\,n!}\,z^n\,{}_2F_1(\alpha+n,\beta+n;\lambda+1+2n;z)\ ,
\end{align*}
with $\alpha=1,\beta=2,\lambda=3/2,z=|x_\perp|^2/|x|^2$ and $G_{2n+3}^{-1}(\upsilon)=\upsilon^{n+3/2}\,{}_2F_1(1+n,2+n;5/2+2n;\upsilon)$.}
\begin{align}\label{eq:3pt bdd Neumann 3 exp}
\begin{aligned}
     \langle\,&\Phi_3^\alpha(x)\, \widehat{\Phi}_{2p}(0)\,\widehat{\Phi}_{2p+1}^{\,\beta}(\infty)\,\rangle 
        \\
            &=f_p\, \delta^{\alpha\beta}\,\frac{|x|}{|x_\perp|^3}\,\left[\frac{N+2}{4}\,G^{-1}_{1}\left(\frac{|x_\perp|^2}{|x|^2}\right)+ 12\,p\,\sum_{n=0}^{\infty}\,\frac{(-1)^n\,(n+1)!}{(n+3/2)_n}\,G^{-1}_{2n+3}\left(\frac{|x_\perp|^2}{|x|^2}\right)\right]   \ . 
\end{aligned}
\end{align}
Compared with the general form \eqref{eq:conformal block expansion main}, the first term is seen as the contribution from $\widehat{\Phi}_1^\alpha$ while the second term indicates an infinite tower of boundary operators with odd integer conformal dimensions, which we denote as $\widehat{\mathsf{O}}_{2n+3}^\alpha$ ($n=0,1,\cdots$), appears in the BOE of $\Phi_3^{\alpha}$:\footnote{Among the tower of the operators, $\widehat{\mathsf{O}}_{3}^\alpha$ can be identified with $\widehat{\Phi}_3^{\alpha}$, but $\widehat{\mathsf{O}}_{2n+3}^\alpha$ are different from $\widehat{\Phi}_{2n+3}^{\,\beta}$ for $n=1,2,\cdots$ since the bulk-boundary two-point function of $\Phi_3^\alpha$ and $\widehat{\Phi}_{2n+3}^{\,\beta}$ vanishes $\langle\,\Phi_3^\alpha\,\widehat{\Phi}_{2n+3}^{\,\beta}\,\rangle=0$ and the BOE of $\Phi_3^\alpha$ does not have $\widehat{\Phi}_{2n+3}^{\,\beta}$.
To be more specific, $\widehat{\mathsf{O}}_{2n+3}^\alpha$ is composed of three $\widehat{\Phi}_1^{\,\alpha}$'s, and $2n$-derivatives w.r.t. parallel directions $\partial/\partial \hat{x}^{a}$ with all parallel indices being contracted. }
\begin{align}\label{eq:btb OPE of phi3 higher-order}
  \Phi_3^{\alpha}(x)
  \supset \frac{b(\Phi_3^\alpha,\widehat{\mathsf{O}}_{2n+3}^\alpha)}{c(\widehat{\mathsf{O}}_{2n+3}^\alpha,\widehat{\mathsf{O}}_{2n+3}^\alpha)}\,|x_\perp|^{2n}\,\widehat{\mathsf{O}}_{2n+3}^\alpha(\hat{x}) \qquad (n=0,1,\cdots)\ .
\end{align}
The ratio $b(\Phi_3^\alpha,\widehat{\mathsf{O}}_{2n+3}^\alpha)/c(\widehat{\mathsf{O}}_{2n+3}^\alpha,\widehat{\mathsf{O}}_{2n+3}^\alpha)$ in the RHS of \eqref{eq:btb OPE of phi3 higher-order} cannot be fixed by the conformal block expansion while we find the following relation by comparing \eqref{eq:3pt bdd Neumann 3 exp} with \eqref{eq:conformal block expansion main}:
\begin{align}    \frac{b(\Phi_3^\alpha,\widehat{\mathsf{O}}_{2n+3}^\alpha)\,c(\widehat{\mathsf{O}}_{2n+3}^\alpha,\widehat{\Phi}_{2p},\widehat{\Phi}_{2p+1}^{\,\beta})}{c(\widehat{\mathsf{O}}_{2n+3}^\alpha,\widehat{\mathsf{O}}_{2n+3}^\alpha)}
        =
            12\,p\,f_p\,\delta^{\alpha\beta} \, \frac{(-1)^n\,(n+1)!}{(n+3/2)_n}\ .\label{eq:b/c 2n+3 Neumann 1}
\end{align}

By repeating the similar discussion for \eqref{eq:3pt bdd Neumann 4}, we obtain
\begin{align}\label{eq:3pt bdd Neumann 4 exp}
\begin{aligned}
    \langle\,&\Phi_3^\alpha(x)\, \widehat{\Phi}_{2p+1}^{\beta}(0)\,\widehat{\Phi}_{2p+2}(\infty)\,\rangle
        \\
        &=g_p\,\delta^{\alpha\beta}\,\frac{|x|}{|x_\perp|^3}
            \,\left[ \frac{N+2}{4}\,G^{-1}_{1}\left(\frac{|x_\perp|^2}{|x|^2}\right)+2\,(N+6p+2)\,\sum_{n=0}^{\infty}\,\frac{(-1)^n\,(n+1)!}{(n+3/2)_n}\,G^{-1}_{2n+3}\left(\frac{|x_\perp|^2}{|x|^2}\right)\right]\ .
\end{aligned}
\end{align}
Compared with \eqref{eq:conformal block expansion main} we find the relation
\begin{align}     \frac{b(\Phi_3^\alpha,\widehat{\mathsf{O}}_{2n+3}^\alpha)\,c(\widehat{\mathsf{O}}_{2n+3}^\alpha,\widehat{\Phi}_{2p+1}^{\beta},\widehat{\Phi}_{2p+2})}{c(\widehat{\mathsf{O}}_{2n+3}^\alpha,\widehat{\mathsf{O}}_{2n+3}^\alpha)}
        =
            2\,(N+6p+2)\,g_p\,\delta^{\alpha\beta}\,\frac{(-1)^n\,(n+1)!}{(n+3/2)_n}\ . \label{eq:b/c 2n+3 Neumann 2}
\end{align}
and the following BOE of $\Phi_3^{\alpha}$:
\begin{align}\label{eq:btb OPE of phi3 all order}
  \Phi_3^{\alpha}(x)
    =
        \frac{1}{|x_\perp|^2}\,\widehat{\Phi}_1^\alpha(\hat{x}) +\sum_{n=0}^{\infty}\,\frac{b(\Phi_3^\alpha,\widehat{\mathsf{O}}_{2n+3}^\alpha)}{c(\widehat{\mathsf{O}}_{2n+3}^\alpha,\widehat{\mathsf{O}}_{2n+3}^\alpha)}\,|x_\perp|^{2n}\,\widehat{\mathsf{O}}_{2n+3}^\alpha(\hat{x})+(\text{descendants})\ .
\end{align}
The ratio $b(\Phi_3^\alpha,\widehat{\mathsf{O}}_{2n+3}^\alpha)/c(\widehat{\mathsf{O}}_{2n+3}^\alpha,\widehat{\mathsf{O}}_{2n+3}^\alpha)$ subject to \eqref{eq:b/c 2n+3 Neumann 1} and \eqref{eq:b/c 2n+3 Neumann 2} will appear again in section \ref{sec:bbb 3pt W1 Neu}.

 \section{Diagrammatic approach} \label{Diagrammatic calculation, Neu}
We perform diagrammatic calculations to derive the leading anomalous dimensions of the boundary composite operators $\widehat{W}_{2p}$ and $\widehat{W}_{2p+1}^{\, \alpha}$ in the O$(N)$ vector model \eqref{eq: action} at the Wilson-Fisher fixed point.\footnote{We employ the minimal subtraction scheme. If we use Pauli-Villars regularization, a mass term must be added to \eqref{eq: action} to preserve the boundary conformal symmetry.}
We consider the theory with the Neumann boundary condition and defer the Dirichlet case to appendix \ref{app: Diagrammatic Calc Dirichlet}. 

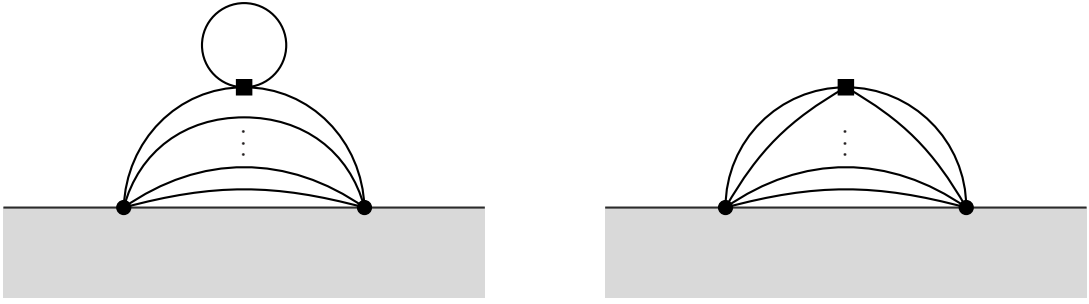
\begin{figure}[t]
    \centering
       \begin{tikzpicture}[transform shape,scale=0.8]
        \begin{scope}[shift={(-5,0)}]        
            \draw[fill=gray!30, draw=none] (-4,0) rectangle (4,-1.5);
            \draw[thick, black!100,opacity=0.8] (-4,0)  -- (4,0);              
            \coordinate (A) at (-2,0) {};
            \coordinate (B) at (2,0) {};
            \coordinate (C) at (0,2) {};  
            \coordinate (D) at (0,1.1) {};
                  
            \filldraw[black,very thick] (A) circle (0.1);
                                           
            \filldraw[black,very thick] (B) circle (0.1);                                        
            \node[rectangle,draw,fill] (r) at (C) {};
            
            \draw[black!100,thick] (A) to[out=15,in=165] (B) ;
            \draw[black!100,thick] (A) to[out=35,in=145] (B) ;
            
            \draw[black!100,thick] (A) to[out=75,in=180] (0,1.5) to[out=0,in=105]  (B) ;
            \draw[black!100,thick] (A) to[out=88,in=180] (C) to[out=0,in=92]  (B) ;

            \draw[thick] (C) arc[
                        start angle=-90,
                        end angle=270,
                        x radius=0.7cm,
                        y radius=0.7cm
                    ] ;

            \node[Black!100,font=\large,rotate around={90:(0,0)}] at (D) {$\cdots$};
      
        \end{scope}

        \begin{scope}[shift={(5,0)}]          
            \draw[fill=gray!30, draw=none] (-4,0) rectangle (4,-1.5);
            \draw[thick, black!100,opacity=0.8] (-4,0)  -- (4,0);    
                        
            \coordinate (A) at (-2,0) {};
            \coordinate (B) at (2,0) {};
            \coordinate (C) at (0,2) {};  
            \coordinate (D) at (0,1.1) {};  
            \coordinate (E) at (0,2) {};             
                  
            \filldraw[black,very thick] (A) circle (0.1);
                                       
            \filldraw[black,very thick] (B) circle (0.1);        
        
            \node[rectangle,draw,fill] (r) at (C) {};

            \draw[black!100,thick] (A) to[out=15,in=165] (B) ;
            \draw[black!100,thick] (A) to[out=35,in=145] (B) ;
            \draw[black!100,thick] (A) to[out=90,in=180] (C) to[out=0,in=90]  (B) ;

            \draw[black!100,thick] (C) to[out=-30,in=120] (B);
            \draw[black!100,thick] (C) to[out=210,in=60] (A);
            
            \node[Black!100,font=\large,rotate around={90:(0,0)}] at (D) {$\cdots$};
        
        \end{scope}

     \end{tikzpicture} 
\caption{One-loop diagrams contributing to the anomalous dimensions of the composite operators $\widehat{W}_{2p}$, $\widehat{W}_{2p+1}^{\, \alpha}$. Black circles ($\bullet$) on the boundary denote boundary operators while black squares ({\tiny $\blacksquare$}) is the bulk $|\Phi_1|^4$ interaction vertex.}
\label{Fig:Quantum correction}
\end{figure}

Let us consider the following two-point functions at the Wilson-Fisher fixed point in $d=(4-\epsilon)$ dimensions:
\begin{align}\label{eq: two-point correlation function, interacting, Neumann}
    \begin{aligned}
         I_{2p}\equiv  \langle\, \widehat{\Phi}_{2p}(\hat{y})\,\widehat{\Phi}_{2p}(0) \,\rangle\, \ ,\qquad  I_{2p+1}^{\, \alpha\beta}\equiv \langle\, \widehat{\Phi}_{2p+1}^{\alpha}(\hat{y})\,\widehat{\Phi}_{2p+1}^{\beta}(0) \,\rangle \, \ ,
    \end{aligned}
\end{align}
where $\widehat{\Phi}_{2p}$ and $\widehat{\Phi}_{2p+1}^{\alpha}$ are bare boundary fields, and the VEV $\langle\, \cdots \, \rangle$ is taken in the interacting vacuum with a boundary.
We can perturbatively calculate \eqref{eq: two-point correlation function, interacting, Neumann} as
\begin{align}
    \begin{aligned}
        I_{2p} &=\, I_{2p, 0} +\delta I_{2p}\, , & \qquad I_{2p, 0}&\equiv  \langle\, \widehat{\Phi}_{2p}(\hat{y})\,\widehat{\Phi}_{2p}(0) \,\rangle_{0} \, \ ,\\
        I_{2p+1}^{\, \alpha\beta} &=\, I_{2p+1, 0}^{\, \alpha\beta} +\delta I_{2p+1}^{\, \alpha\beta}\, , & \qquad I_{2p+1, 0}^{\, \alpha\beta} &\equiv  \langle\, \widehat{\Phi}_{2p+1}^{\alpha}(\hat{y})\,\widehat{\Phi}_{2p+1}^{\beta}(0) \,\rangle_{0} \, \ ,
    \end{aligned}
\end{align}
where $\langle\, \cdots \, \rangle_0$ stands for the VEV in the vacuum of the $(4-\epsilon)$-dimensional free theory with a boundary.
$\delta I_{2p}$ and $\delta I_{2p+1}^{\alpha\beta}$ are the quantum corrections to the free propagators, whose diagram is shown in figure \ref{Fig:Quantum correction}. 

These corrections diverge and we need to introduce the renormalized boundary fields by
\begin{align}\label{eq:wave function ren Neumann}
    \begin{aligned}
        \widehat{W}_{2p}=Z_{2p}^{-1}\,\widehat{\Phi}_{2p} \ ,\qquad \widehat{W}_{2p+1}^{\,\alpha}=Z_{2p+1}^{-1}\, \widehat{\Phi}_{2p+1}^{\alpha}\  , 
    \end{aligned}
\end{align}
where $Z_{2p}$ and $Z_{2p+1}$ are the wave-function renormalization factors.
Note that the two-point function of $\widehat{W}$ differs from that of the free fields by order $O(\epsilon)$.
We choose $Z_{2p}$ and $Z_{2p+1}$ so as to cancel the divergence appearing in $\delta I_{2p}$ and $\delta I_{2p+1}^{\, \alpha\beta}$.
In the minimal subtraction scheme, $Z_{2p}$ and $Z_{2p+1}$ can be expanded as
\begin{align}\label{eq:wave function ren factor Neumann}
    \begin{aligned}
        Z_{2p} &=1+\delta Z_{2p}\, , \qquad
        Z_{2p+1} &=1+\delta Z_{2p+1}\, , 
    \end{aligned}
\end{align}
$\delta Z_{2p}$ and $\delta Z_{2p+1}$ are of order $O(\lambda)$ and related to the conformal dimensions of $\widehat{W}_{2p}$ and $\widehat{W}_{2p+1}^{\, \alpha}$ by
\begin{align}\label{eq: relation between anomalous dimensions and wavefunction renormalization}
  \widehat{\Delta}_n =n\, \frac{d-2}{2}+\widehat{\gamma}_{n} \  , \qquad  \widehat{\gamma}_{n}&\equiv\beta_{\lambda} \left. \frac{\text{d}\,  \ln Z_{n}}{\text{d} \lambda}\right|_{\lambda=\lambda_\ast}\, , 
\end{align}
where $\beta_{\lambda}$ and $\lambda_\ast$ are given in \eqref{eq:beta function} and \eqref{eq: Wilson-Fisher FP}, respectively.
The anomalous dimension $\widehat{\gamma}_{n}$ has a power expansion in $\epsilon$:
\begin{align}\label{eq: relation between anomalous dimensions and wavefunction renormalization first}
\widehat{\gamma}_{n}=\widehat{\gamma}_{n,1}\,\epsilon+\widehat{\gamma}_{n,2}\,\epsilon^2+\cdots\ ,\qquad \widehat{\gamma}_{n,1}&=-\lambda \left. \frac{\text{d}\,  \delta Z_{n}}{\text{d} \lambda}\right|_{\lambda=\lambda_\ast}\, .
\end{align}
We focus on the leading part $\widehat{\gamma}_{n,1}$ and evaluate $I_{2p}$ at one-loop level.
By a standard perturbative calculation \cite{Wilson:1973jj, Peskin:1995ev, Collins:1984xc, Kleinert:2001ax, kardar_2007}, $\delta I_{2p}$ becomes\footnote{We use the following integral formula:
\begin{align*}
    \int_{\mathbb{R}^{d}_{+}} \d^d x\, \frac{1}{x_\perp ^{2\alpha}\, |x|^{2\beta}\, |x-\hat{y}|^{2\gamma}}= \frac{F_{\alpha, \beta, \gamma}}{|\hat{y}|^{2\alpha+2\beta+2\gamma-d}} \, ,
\end{align*}
where
\begin{align*}
    F_{\alpha, \beta, \gamma}=\frac{\Gamma(\frac{1}{2}-\alpha)\, \Gamma(\alpha+\beta+\gamma-\frac{d}{2})\, \Gamma(\frac{d}{2}-\alpha-\gamma)\, \Gamma(\frac{d}{2}-\alpha-\beta)\, \pi^{\frac{d-1}{2}}}{2\, \Gamma(\beta)\, \Gamma(\gamma)\, \Gamma(d-2\alpha-\beta-\gamma)}\ .
\end{align*}}
\begin{align}\label{eq:delta I_2p}
    \begin{aligned}
        \delta I_{2p} &= -\frac{\lambda_{0}}{4!}\int _{\mathbb{R}^{d}_{+}} \d^{d}x \, \left\langle \, |\Phi_{1}|^4 (x)\, \widehat{\Phi}_{2p}(\hat{y})\,  \widehat{\Phi}_{2p}(0)\right\rangle_0 \\
        &= -\frac{N+8p-6}{\epsilon}\, \pi^2 \lambda_{0}\, I_{2p, 0}+\frac{4\, p}{|\hat{y}|^{2}}\, \delta I_{2p-1}^{\alpha\alpha} +O(\lambda_{0}^2\, , \, \epsilon^0)\, ,\\
    \end{aligned}
\end{align}
where we used the correlation functions in the $(4-\epsilon)$-dimensional free theory in section \ref{eq:Correlation functions in arbitrary dimensions Neu}.
Furthermore, by using \eqref{eq:wave function ren Neumann} and \eqref{eq:wave function ren factor Neumann}, \eqref{eq:delta I_2p} reduces to the following recursion relation between $\delta Z_{2p}$ and $\delta Z_{2p-1}$:
\begin{align}\label{eq: recursion relation between Z_2p and Z_{2p-1}}
    \delta Z_{2p}-\delta Z_{2p-1}=-\frac{N+8p-6}{2\,\epsilon}\, \pi^2\, \lambda+O(\lambda^{2}, \epsilon^0)\, .
\end{align}
In the same way as $I_{2p}$, $I_{2p-1}^{\, \alpha\beta}$ can be calculated at one-loop level as
    \begin{align}\label{eq:deltaI_2p-1}
    \delta I^{\alpha\beta}_{2p-1}=-\frac{24p-N-26}{3\,\epsilon}\, \pi^2 \lambda_{0}\,I^{\alpha\beta}_{2p-1, 0}+\frac{2\, (N+2p-2)\, \delta^{\alpha\beta}}{N\, |\hat{y}|^{2}} \delta I_{2p-2}\, +O(\lambda_{0}^2\, , \, \epsilon^0) \ . 
    \end{align}
It follows from \eqref{eq:wave function ren Neumann} and \eqref{eq:wave function ren factor Neumann} that \eqref{eq:deltaI_2p-1} yields 
\begin{align}\label{eq: recursion relation between Z_{2p-1} and Z_{2p-2}}
    \delta Z_{2p-1}-\delta Z_{2p-2}=-\frac{24p-N-26}{6\,\epsilon}\, \pi^2\, \lambda +O(\lambda^2, \epsilon^0)\, .
\end{align}
By solving the above two recursion relations \eqref{eq: recursion relation between Z_2p and Z_{2p-1}} and \eqref{eq: recursion relation between Z_{2p-1} and Z_{2p-2}} under the initial condition $\delta Z_{0}=0$, the wave-function renormalization factors are determined in the minimal subtraction scheme:
\begin{align}\label{wave func ren: Neumann, diagram}
\begin{aligned}
   \delta Z_{2p}&=-\frac{p\, (N+12p-10)}{3\,\epsilon}\, \pi^2\lambda\, +O(\lambda^2)\, , \\
    \delta Z_{2p+1}&=-\frac{(2p-1)\,N+2\,(12p^2+2p-1)}{6\,\epsilon}\,\pi^2\lambda +O(\lambda^2) \, .
\end{aligned}
\end{align}
By substituting \eqref{wave func ren: Neumann, diagram} to \eqref{eq: relation between anomalous dimensions and wavefunction renormalization first}, we obtain the leading anomalous dimensions of $\widehat{W}_{2p}$ and $\widehat{W}_{2p+1}^{\, \alpha}$:
\begin{align}\label{eq: Diagrammatic result, Neumann}
        \widehat{\gamma}_{2p,1}=\frac{p\, (N+12p-10)}{N+8} \ ,\qquad
        \widehat{\gamma}_{2p+1,1}= \frac{(2p-1)\,N+2\,(12p^2+2p-1)}{2\,(N+8)}\ .
\end{align}
Notice that $\widehat{\gamma}_{1,1}=-(N+2)/2(N+8)$ and $\widehat{\gamma}_{2,1}=(N+2)/(N+8)$ agree with the known results given in \cite[section 3]{McAvity:1995zd}.

\section{Axiomatic approach} \label{sec:Rychkov-Tan Axioms}
The goal of this section is to explore the critical behavior of the O$(N)$ model with the Neumann boundary condition from the axiomatic method of \cite{Rychkov:2015naa}.
We postulate the following three axioms under which we reproduce the diagrammatic results in the last section \eqref{eq: Diagrammatic result, Neumann}.
\begin{itemize}\setlength{\leftskip}{8mm}
    \begin{shaded}
    \item[\textbf{Axiom \mylabel{bcftaxiom1}{\RomanNumeralCaps{1}}.}] In the presence of a boundary, the theory at the Wilson-Fisher fixed point has the \textbf{boundary conformal symmetry}.
    \end{shaded}
\end{itemize}
  \begin{itemize}\setlength{\leftskip}{10mm}
  \begin{shaded}
      \item[\textbf{Axiom \mylabel{bcftaxiom2}{\RomanNumeralCaps{2}}.}] \textbf{For a bulk/boundary local operator} $\CO_{\text{free}}$/$\widehat{\CO}_{\text{free}}$ in the free theory with a boundary, there exists a local operator at the Wilson-Fisher fixed point denoted by $\CO_{\text{WF}}$/$\widehat{\CO}_{\text{WF}}$, which tends to $\CO_{\text{free}}$/ $\widehat{\CO}_{\text{free}}$ in the limit $\epsilon \to 0$.
  \end{shaded}
\end{itemize}
\begin{itemize}\setlength{\leftskip}{12mm}
\begin{shaded}
    \item[\textbf{Axiom \mylabel{dcftaxiom3}{\RomanNumeralCaps{3}}.}] At the Wilson-Fisher fixed point, \textbf{two bulk operators} $W_1^\alpha$ and $W_3^\alpha$, which tend to $\Phi_1^\alpha$ and $\Phi_3^\alpha$ as $\epsilon\to0$, are related by the following equation of motion:
    \begin{align}\label{eq:classical EoM O(N)}
    \Box_x \, W_1^\alpha(x)=\kappa\,W_3^\alpha(x)\  ,
\end{align}
where $\Box_x$ is the Laplacian in $d=(4-\epsilon)$ dimensions.
\end{shaded}
\end{itemize}
We emphasize that the parameters $\kappa$ and $\Delta_1$ are fixed by the bulk criticality \cite{Rychkov:2015naa}:
\begin{align}\label{eq:bulk scalar O(N) zeroth order}
    \begin{aligned}
        \kappa=\frac{2}{N+8}\,\epsilon+O(\epsilon^2)\ , \qquad  \Delta_1=1-\frac{1}{2}\,\epsilon+O(\epsilon^2) \ .
    \end{aligned}
\end{align}
See \cite{Nishioka:2022qmj,Yamaguchi:2016pbj} for more details.

\subsection{Lowest-lying boundary local operator}\label{sec:Neu lowest RT}
We begin with the lowest-lying boundary local operator $\widehat{W}_1^\alpha$ and derive its conformal dimension up to $\epsilon$ along the line of \cite{Yamaguchi:2016pbj}.

From the conformal symmetry, one has the following BOE of $W_1^\alpha$:
\begin{align}\label{eq:W1_bOPE_Neumann}
    W_1^\alpha(x)
        \supset
             D\,\frac{1}{|x_\perp|^{\Delta_1-\widehat{\Delta}_{1}}}\,\widehat{W}_{1}^\alpha(\hat{x}) \ .
\end{align}
For this BOE to be compatible with \eqref{eq:OPE of phi1 O(N) Neumann} in the free theory, the coefficient $D$ and the conformal dimension $\widehat{\Delta}_{1}$ should be
\begin{align}
    D = 1+O(\epsilon)\ ,\qquad \widehat{\Delta}_{1}=1+O(\epsilon) \ .
\end{align}
We substitute the BOE \eqref{eq:W1_bOPE_Neumann} to the equation of motion \eqref{eq:classical EoM O(N)} and end up with\footnote{The $d$-dimensional Laplacian $\Box$ in the presence of a $p$-dimensional planar defect acts as
\begin{align}
    \Box \left( |x|^{-a}\,|x_\perp|^{-b}\right) 
        =
            a\,(a + 2b + 2 - d)\,|x|^{-a-2}\,|x_\perp|^{-b} 
                +
                b\,(b + 2 + p - d)\,|x|^{-a}\,|x_\perp|^{-b-2} \ .
\end{align}
}
\begin{align}
\begin{aligned}
     W_3^\alpha(x)
        &=
            \frac{1}{\kappa}\,\Box\,W_1^\alpha(x)\\
        &\supset
    \frac{D}{\kappa}\,\frac{(\Delta_1-\widehat{\Delta}_{1})(\Delta_1-\widehat{\Delta}_{1}+1)}{|x_\perp|^{\Delta_1-\widehat{\Delta}_{1}+2}}\,\widehat{W}_{1}^\alpha(\hat{x})\ .     
\end{aligned}
\end{align}
This relation should reduce to \eqref{eq:OPE of phi3 O(N) Neumann} in the $\epsilon \to 0$ limit, hence we find
\begin{align}
    \frac{N+2}{4}
        =
            \frac{(\Delta_1 - \widehat{\Delta}_{1})(\Delta_1 - \widehat{\Delta}_{1}+1)}{\kappa} + O(\epsilon) \ .
\end{align}
By plugging $\kappa$ and $\Delta_1$ in \eqref{eq:bulk scalar O(N) zeroth order} into the above relation, the dimension is given by
\begin{align}\label{eq:anomalous dim lowest Neu}
    \begin{aligned}
           \widehat{\Delta}_{1}
                &=
               \frac{d-2}{2}-\frac{N+2}{2\,(N+8)}\,\epsilon+O(\epsilon^2) \\
               &=1-\frac{N+5}{N+8}\,\epsilon+O(\epsilon^2)\ , 
    \end{aligned}
\end{align}
which agrees with the known result in literature \cite[equation (3.10)]{McAvity:1995zd}.

\subsection{Boundary composite operators}\label{sec:Neu higher order}
We proceed to derive the conformal dimensions $\widehat{\Delta}_{2p}$ and $\widehat{\Delta}_{2p+1}$ of the boundary local operators $\widehat{W}_{2p}, \widehat{W}_{2p+1}^{\,\alpha}$
which reduce to $\widehat{\Phi}_{2p},\widehat{\Phi}_{2p+1}^{\,\alpha}$ in the free limit, respectively.

We will determine the leading anomalous dimensions of these composite operators $\widehat{\gamma}_{n,1}$ \eqref{eq: relation between anomalous dimensions and wavefunction renormalization}.
These composite operators do not appear in the BOE of the free bulk scalar $\Phi_1^\alpha$, hence we cannot employ the strategy used in the last section.
Instead, we will take the following steps in this section:
\begin{itemize}
    \item 
        We implement the equation of motion \eqref{eq:classical EoM O(N)} to calculate the BOE of $W_1$ at order $O(\epsilon)$. (section \ref{sec:btd operator product expansion W1 Neu})
    \item By using the BOE, we evaluate the following bulk-boundary-boundary three-point functions at order $O(\epsilon)$ (section \ref{sec:bbb 3pt W1 Neu}):
        \begin{align}\label{eq:bulk-bdy-bdy 3pt func}
            \langle\,W_1^\alpha(x)\,\widehat{W}_{2p}(0)\,\widehat{W}_{2p+1}^{\,\beta}(\infty)\,\rangle\ , \qquad
            \langle\,W_1^\alpha(x)\,\widehat{W}_{2p+1}^{\,\beta}(0)\,\widehat{W}_{2p+2}(\infty)\,\rangle \ .
        \end{align}
    \item 
        It turns out that these two correlators \eqref{eq:bulk-bdy-bdy 3pt func} have unphysical singularities at $|\hat{x}|=0$.
        Resolving these singularities imposes some constraints on the anomalous dimensions, whose solutions completely match with the diagrammatic results \eqref{eq: Diagrammatic result, Neumann}. (section \ref{sec:Constraint from analyticity})
\end{itemize}

\subsubsection{Boundary operator expansion of $W_1^\alpha$}\label{sec:btd operator product expansion W1 Neu}
In the free theory, $W_1^\alpha$ reduces to $\Phi_1^\alpha$ and its BOE only contains $\widehat{\Phi}_1^\alpha$ (see section \ref{sec:Boundary operator expansion of Phi1}).
However, at order $O(\epsilon)$, the other operators start to contribute to the BOE.
More specifically, $W_1^\alpha$ couples to a series of operators $\widehat{\mathsf{O}}_{2n+3}^{\prime\,\alpha}$ ($n\in\mathbb{Z}_{\geq0}$) that has the conformal dimension $\widehat{\Delta}'_{2n+3}=2n+3+O(\epsilon)$ and can be identified with $\widehat{\mathsf{O}}_{2n+3}^\alpha$ when $\epsilon=0$:
\begin{align}
\lim_{\epsilon\to0}\,\widehat{\mathsf{O}}_{2n+3}^{\prime\,\alpha} (\hat{x}) =\widehat{\mathsf{O}}_{2n+3}^\alpha (\hat{x}) \ .
\end{align}
Namely, the BOE takes the form:
\begin{align}\label{eq:OPE of W1 all order}
    \begin{aligned}
        W_1^\alpha(x)
            &=
                \frac{D}{|x_\perp|^{\Delta_1-\widehat{\Delta}_1}}\,\widehat{W}_1^{\,\alpha}(\hat{x})\\
            &\qquad
                +
   \sum_{n=0}^{\infty}\,\frac{b(W_1^\alpha,\widehat{\mathsf{O}}_{2n+3}^{\prime\,\alpha})/c(\widehat{\mathsf{O}}_{2n+3}^{\prime\,\alpha},\widehat{\mathsf{O}}_{2n+3}^{\prime\,\alpha})}{|x_\perp|^{\Delta_1-\widehat{\Delta}'_{2n+3}}}\,\widehat{\mathsf{O}}_{2n+3}^{\prime\,\alpha}(\hat{x})+(\text{descendants})\ ,
    \end{aligned}
\end{align}
with $D=1+O(\epsilon)$ being introduced in section \ref{sec:Neu lowest RT}.
The $O(\epsilon)$ coefficients $b(W_1^\alpha,\widehat{\mathsf{O}}_{2n+3}^{\prime\,\alpha})/c(\widehat{\mathsf{O}}_{2n+3}^{\prime\,\alpha},\widehat{\mathsf{O}}_{2n+3}^{\prime\,\alpha})$ can be evaluated by using the equation of motion \eqref{eq:classical EoM O(N)}.
Acting the Laplacian on the LHS of \eqref{eq:OPE of W1 all order} leads
\begin{align}
    \begin{aligned}
        W_3^\alpha(x)
            &\supset
                \sum_{n=0}^{\infty}\,\frac{b(W_1^\alpha,\widehat{\mathsf{O}}_{2n+3}^{\prime\,\alpha})}{\kappa\,c(\widehat{\mathsf{O}}_{2n+3}^{\prime\,\alpha},\widehat{\mathsf{O}}_{2n+3}^{\prime\,\alpha})}\,\frac{(\Delta_1-\widehat{\Delta}'_{2n+3})(\Delta_1-\widehat{\Delta}'_{2n+3}+1)}{|x_\perp|^{\Delta_1-\widehat{\Delta}'_{2n+3}+2}}\,\widehat{\mathsf{O}}_{2n+3}^{\prime\,\alpha}(\hat{x}) \ .
    \end{aligned}
\end{align}
Comparing this BOE with \eqref{eq:btb OPE of phi3 all order}, we find
\begin{align}\label{eq:W1 to higher order OPE coeff}
    \frac{b(W_1^\alpha,\widehat{\mathsf{O}}_{2n+3}^{\prime\,\alpha})}{c(\widehat{\mathsf{O}}_{2n+3}^{\prime\,\alpha},\widehat{\mathsf{O}}_{2n+3}^{\prime\,\alpha})}
        =
            \frac{\kappa}{2\,(n+1)(2n+1)}\frac{b(\Phi_3^\alpha,\widehat{\mathsf{O}}_{2n+3}^\alpha)}{c(\widehat{\mathsf{O}}_{2n+3}^\alpha,\widehat{\mathsf{O}}_{2n+3}^\alpha)}+O(\epsilon^2)\ .
\end{align}

\subsubsection{Bulk-boundary-boundary three-point functions involving $W_1^\alpha$}\label{sec:bbb 3pt W1 Neu}
We are now in a position to calculate the bulk-boundary-boundary three-point functions involving $W_1^\alpha$.
Using \eqref{eq:OPE of W1 all order}, the conformal block expansion of $\langle\,W_{1}^\alpha\, \widehat{W}_{2p}\,\widehat{W}_{2p+1}^{\,\beta}\,\rangle$ becomes
\begin{align}
    \begin{aligned}
    \langle\,& W_1^\alpha(x)\, \widehat{W}_{2p}(0)\,\widehat{W}_{2p+1}^{\,\beta}(\infty)\,\rangle
        =
            \frac{1}{|x_\perp|^{\Delta_1}\,|x|^{\widehat{\Delta}_{2p}-\widehat{\Delta}_{2p+1}}}\\
        &\qquad\cdot
            \left[ D\cdot c(\widehat{W}_1^{\,\alpha},\widehat{W}_{2p},\widehat{W}_{2p+1}^{\,\beta})\,G^{\widehat{\Delta}_{2p}-\widehat{\Delta}_{2p+1}}_{\widehat{\Delta}_1}\left(\frac{|x_\perp|^2}{|x|^2}\right)\right.\\
        &\qquad\qquad\left.     
            +
            \sum_{n=0}^{\infty}\,\frac{b(W_1^\alpha,\widehat{\mathsf{O}}_{2n+3}^{\prime\,\alpha})\,c(\widehat{\mathsf{O}}_{2n+3}^{\prime\,\alpha},\widehat{W}_{2p+1}^{\,\beta},\widehat{W}_{2p+2})}{c(\widehat{\mathsf{O}}_{2n+3}^{\prime\,\alpha},\widehat{\mathsf{O}}_{2n+3}^{\prime\,\alpha})}\,G^{\widehat{\Delta}_{2p}-\widehat{\Delta}_{2p+1}}_{\widehat{\Delta}'_{2n+3}}\left(\frac{|x_\perp|^2}{|x|^2}\right)\right]\ .
    \end{aligned}
\end{align}
The first term in the parenthesis can be evaluated as
\begin{align}
    \begin{aligned}
        G^{\widehat{\Delta}_{2p}-\widehat{\Delta}_{2p+1}}_{\widehat{\Delta}_1}(\upsilon)&=\upsilon^{\widehat{\Delta}_1/2}\,{}_2F_1\left(\frac{\widehat\gamma_{1,1}+\widehat\gamma_{2p,1}-\widehat\gamma_{2p+1,1}}{2}\,\epsilon,1;1/2;\upsilon\right)+O(\epsilon^2)\\
            &=
                \upsilon^{\widehat{\Delta}_1}+(\widehat\gamma_{1,1}+\widehat\gamma_{2p,1}-\widehat\gamma_{2p+1,1})\,\epsilon\,\upsilon^{3/2}\,{}_2F_1(1,1;3/2;\upsilon)+O(\epsilon^2)\ .
    \end{aligned}
\end{align}
On the other hand, for the remaining terms, the coefficients are already of order $O(\epsilon)$ \eqref{eq:W1 to higher order OPE coeff} and we have
\begin{align}
    \begin{aligned}
    & \sum_{n=0}^{\infty}\,\frac{b(W_1^\alpha,\widehat{\mathsf{O}}_{2n+3}^{\prime\,\alpha})\,c(\widehat{\mathsf{O}}_{2n+3}^{\prime\,\alpha},\widehat{W}_{2p+1}^{\,\beta},\widehat{W}_{2p+2})}{c(\widehat{\mathsf{O}}_{2n+3}^{\prime\,\alpha},\widehat{\mathsf{O}}_{2n+3}^{\prime\,\alpha})}\,G^{\widehat{\Delta}_{2p}-\widehat{\Delta}_{2p+1}}_{\widehat{\Delta}'_{2n+3}}(\upsilon)\\
       &\quad 
        =
           6\,\kappa \,p\,f_p\, \delta^{\alpha\beta} \, \upsilon^{3/2}\,{}_2F_1(1,1;3/2;\upsilon)+O(\epsilon^2)\ ,
    \end{aligned}
\end{align}
where we used \eqref{eq:b/c 2n+3 Neumann 1} and the sum rule for hypergeometric functions that can be verified by expanding in powers of $z$ and comparing both sides order by order
\begin{align}
    {}_2F_1(1,1;3/2;z)=\sum_{n=0}^{\infty}\,\frac{(-1)^n\,n!}{(2n+1)\,(n+3/2)_n} \, z^{n}\,{}_2F_1(n+1,n+2;2n+5/2;z)\ .
\end{align}
Combining all the calculations, the three-point function $\langle\, W_1^\alpha\, \widehat{W}_{2p}\,\widehat{W}_{2p+1}^{\,\beta}\,\rangle$ simplifies to\footnote{We used
\begin{align}
    c(\widehat{W}_1^{\,\alpha},\widehat{W}_{2p},\widehat{W}_{2p+1}^{\,\beta})
        =
            c(\widehat{\Phi}_1^{\,\alpha},\widehat{\Phi}_{2p},\widehat{\Phi}_{2p+1}^{\,\beta})+O(\epsilon)
        =
            f_p\,\delta^{\alpha\beta}+O(\epsilon)\ .
\end{align}
}
\begin{align}\label{eq:W1 3pt expansion 1}
    \begin{aligned}
    \langle\, &W_1^\alpha(x)\, \widehat{W}_{2p}(0)\,\widehat{W}_{2p+1}^{\,\beta}(\infty)\,\rangle\\
        &=
            c(\widehat{W}_1^{\,\alpha},\widehat{W}_{2p},\widehat{W}_{2p+1}^{\,\beta})\, \frac{D}{|x_\perp|^{\Delta_1-\widehat{\Delta}_1}\,|x|^{\widehat{\Delta}_1+\widehat{\Delta}_{2p}-\widehat{\Delta}_{2p+1}}}  \\
        &\qquad 
            +
                f_p\, \delta^{\alpha\beta}\,\bigg[ \left(\widehat\gamma_{1,1}+\widehat\gamma_{2p,1}-\widehat\gamma_{2p+1,1}\right)\,\epsilon+ 6\,\kappa\,p \bigg]\,\frac{|x_\perp|^2}{|x|^2} \, {}_2F_1\left(1,1;\frac{3}{2};\frac{|x_\perp|^2}{|x|^2}\right)+O(\epsilon^2)\ .
    \end{aligned}
\end{align}

The other correlator of our interest $\langle\, W_1^\alpha\, \widehat{W}_{2p+1}^{\,\beta}\,\widehat{W}_{2p+2}\,\rangle$ can be derived in a similar manner:
\begin{align}\label{eq:W1 3pt expansion 2}
    \begin{aligned}
    \langle\,& W_1^\alpha(x)\, \widehat{W}_{2p+1}^{\,\beta}(0)\,\widehat{W}_{2p+2}(\infty)\,\rangle\\
        &=
            c(\widehat{W}_1^{\,\alpha},\widehat{W}_{2p}^{\,\beta},\widehat{W}_{2p+2})\, \frac{D}{|x_\perp|^{\Delta_1-\widehat{\Delta}_1}\,|x|^{\widehat{\Delta}_1+\widehat{\Delta}_{2p+1}-\widehat{\Delta}_{2p+2}}}\\
        & ~+
            g_p\,\delta^{\alpha\beta}\,\bigg[ \left(\widehat\gamma_{1,1}+\widehat\gamma_{2p+1,1}-\widehat\gamma_{2p+2,1}\right)\,\epsilon+ \kappa\,(N+6p+2)\bigg]\, \frac{|x_\perp|^2}{|x|^2} \, {}_2F_1\left(1,1;\frac{3}{2};\frac{|x_\perp|^2}{|x|^2}\right)+O(\epsilon^2)\ .
    \end{aligned}
\end{align}

\subsubsection{Constraints from analyticity}\label{sec:Constraint from analyticity}
From the asymptotic behavior of the hypergeometric function\footnote{Recall Kummer's connection formula for hypergeometric functions;
\begin{align*}
    \begin{aligned}
    {}_2F_1(\alpha,\beta;\gamma;z)&=\frac{\Gamma(\gamma)\Gamma(\gamma-\alpha-\beta)}{\Gamma(\gamma-\alpha)\Gamma(\gamma-\beta)}\,{}_2F_1(\alpha,\beta;\alpha+\beta-\gamma+1;1-z)\\
    &\qquad +\frac{\Gamma(\gamma)\Gamma(\alpha+\beta-\gamma)}{\Gamma(\alpha)\Gamma(\beta)}\,(1-z)^{\gamma-\alpha-\beta}\,{}_2F_1(\gamma-\alpha,\gamma-\beta;\gamma-\alpha-\beta+1;1-z)\ .
    \end{aligned}
\end{align*}
}
\begin{align}
    {}_2F_1\left(1,1;\frac{3}{2};\frac{|x_\perp|^2}{|x|^2}\right)\xrightarrow[|\hat{x}|\sim 0]{} \frac{\pi}{2}\cdot \frac{|x_\perp|}{|\hat{x}|}+(\text{less singular terms})\ ,
\end{align}
we find that both \eqref{eq:W1 3pt expansion 1} and \eqref{eq:W1 3pt expansion 2} are singular at $|\hat{x}|=0$, where the bulk operator has a finite separation from the boundary and any pair of the operators collide.
Hence, for the correlators to be holomorphic for all non-coincident configurations, we must require coefficients in front of ${}_2F_1(1,1;3/2;|x_\perp|^2/|x|^2)$ to vanish:
\begin{align}\label{eq:Neu req from regularity}
    \begin{aligned}
    (\widehat\gamma_{1,1}+\widehat\gamma_{2p,1}-\widehat\gamma_{2p+1,1})\,\epsilon+ 6\,\kappa\,p 
        &=
            0+O(\epsilon^2)\ ,\\
    ( \widehat\gamma_{1,1}+\widehat\gamma_{2p+1,1}-\widehat\gamma_{2p+2,1})\,\epsilon+ \kappa\,(N+6p+2)
        &=
            0+O(\epsilon^2)\ .
    \end{aligned}
\end{align}
These constraints lead to the recursion relations:
\begin{align}\label{eq:Neu recursion rel}
    \widehat{\gamma}_{2p+1,1}
        =
            \widehat{\gamma}_{2p,1}-\frac{N-24p+2}{2\,(N+8)}\ ,\qquad 
    \widehat{\gamma}_{2p+2,1}
        =
            \widehat{\gamma}_{2p+1,1}+\frac{3\,(N+8p+2)}{2\,(N+8)}\ .
\end{align}
It is straightforward to see that the solutions to these recursion relations correctly reproduce the diagrammatic results \eqref{eq: Diagrammatic result, Neumann}.

\acknowledgments
We are grateful to T.\,Onogi for valuable discussions.
The work of T.\,N. was supported in part by the JSPS Grant-in-Aid for Scientific Research (C) No.19K03863, Grant-in-Aid for Scientific Research (A) No.\,21H04469, and
Grant-in-Aid for Transformative Research Areas (A) ``Extreme Universe''
No.\,21H05182 and No.\,21H05190.
The work of Y.\,O. was supported by Forefront Physics and Mathematics Program to Drive Transformation (FoPM), a World-leading Innovative Graduate Study (WINGS) Program, the University of Tokyo.
The work of Y.\,O. was also supported by JSPS fellowship for young students, MEXT, and by JSR fellowship, the University of Tokyo.

\appendix

\section{Dirichlet boundary condition}\label{app:Dirichlet boundary condition}
We deal with the model \eqref{eq: action} with the Dirichlet boundary condition.
We will perform a similar analysis to the Neumann case below.

\subsection{The free $\textrm{O}(N)$ model with Dirichlet boundary condition}\label{app:Correlation functions in free theories in the Dirichlet case}
Under the Dirichlet boundary condition
\begin{align}\label{eq:Dirichlet boundary condition for Phi1}
     \Phi_1^\alpha(x)\big|_{x_\perp=0}=0 \ ,
\end{align}
the lowest-lying boundary local operator in free theory is defined by
\begin{align}\label{eq:Dirichlet boundary lowest Phi}
     \widehat{\Psi}_2^\alpha(\hat{x})\equiv \lim_{x_\perp\to0}\,x_\perp^{-1}\,\Phi_1^\alpha(x)\ .
\end{align}
There are also boundary composite operators with even integer conformal dimensions:
 \begin{align}\label{eq:free limit Dirichlet}
                \widehat{\Psi}_{4p}(\hat{x})\equiv\lim_{x_\perp\to0}\, x_{\perp}^{-2p}\, |\Phi_1|^{2p}(x)\,  ,\qquad \widehat{\Psi}_{4p+2}^{\,\alpha}(\hat{x})\equiv\lim_{x_\perp\to0}\, x_{\perp}^{-2p-1}\, \Phi_1 ^\alpha |\Phi_1|^{2p}(x)\ .
\end{align}

\subsubsection{Correlation functions in arbitrary dimensions}
We summarize the correlation functions for free scalar fields subject to the Dirichlet boundary condition in $d$ dimensions.

The bulk two-point function satisfies the same differential equation as the Neumann case \eqref{eq:4d free scalar propagator EoM}. However, the solution is different:
\begin{align}\label{eq:4d free scalar propagator Dirichlet}
 \langle\,\Phi_{1}^\alpha(x_1)\,\Phi_{1}^\beta(x_2)\,\rangle=\delta^{\alpha\beta}\,\left[\frac{1}{|x_1-x_2|^{d-2}}- \frac{1}{|x_1-\bar{x}_2|^{d-2}}\right]\ .
\end{align}
Then, we have
\begin{align}\label{eq:Dirichlet composite bulk 1pt}
  \langle\,\Phi_{1}^\alpha\Phi_{1}^\beta(x)\,\rangle=-\frac{\delta^{\alpha\beta}}{2^{d-2}\,|x_\perp|^{d-2}}\ ,\qquad   \langle\,|\Phi_1|^{\,2}(x)\,\rangle=-\frac{N}{2^{d-2}\,|x_\perp|^{d-2}}\ .
\end{align}
The two-point functions involving the lowest boundary local operator $\widehat{\Psi}_{2}^{\,\alpha}$ defined in \eqref{eq:Dirichlet boundary lowest Phi} are
\begin{align}\label{eq:2pt Phi Dir}
    \langle\,\Phi_1^\alpha(x)\,\widehat{\Psi}_{2}^{\,\beta}(\hat{y})\,\rangle
        =
            \frac{ 2(d-2)\,  \delta^{\alpha\beta}\,x_{\perp}}{|x-\hat{y}|^d}\ ,\qquad
       \langle\,\widehat{\Psi}_{2}^{\,\alpha}(\hat{y}_1)\,\widehat{\Psi}_{2}^{\,\beta}(\hat{y}_2)\,\rangle
        =
            \frac{2(d-2)\, \delta^{\alpha\beta}}{|\hat{y}_{12}|^d} \ .
\end{align}
Any correlators can be calculated by applying Wick's theorem.
For instance,
\begin{align}\label{eq:Dir btb 2pt Phi3}
\begin{aligned}
      \langle\,\Phi_3^\alpha(x)\,\widehat{\Psi}_{2}^{\,\beta}(\hat{y})\,\rangle
        &=
            -\frac{(d-2)(N/2+1)\,\delta^{\alpha\beta}}{2^{d-4}\,|x-\hat{y}|^{d}\,|x_\perp|^{d-3}}\ ,\\
            \langle\,\Phi_3^\alpha(x)\,\widehat{\Psi}_{6}^{\,\beta}(\hat{y})\,\rangle
                &=
                \frac{  32(d-2)^3\,(N/2+1)\,\delta^{\alpha\beta}\,x_\perp^3}{|x-\hat{y}|^{3d}}\ ,\\
   \langle\,\widehat{\Psi}_{6}^\alpha(\hat{y}_1)\,\widehat{\Psi}_{6}^{\,\beta}(\hat{y}_2)\,\rangle
                        &=
                        \frac{  32(d-2)^3\,(N/2+1)\,\delta^{\alpha\beta}}{|\hat{y}_{12}|^{3d}}\ .
\end{aligned}
\end{align}
We record other correlators which are necessary for the rest of this appendix.
\paragraph{Boundary two-point functions.}
\begin{align}
            \langle\, \widehat{\Psi}_{4p}(\hat{y}_1)\,\widehat{\Psi}_{4p}(\hat{y}_2) \,\rangle = \frac{N\, b_{p-1}}{|\hat{y}_{12}|^{2\, p\, d}} \, ,\qquad
             \langle\,  \widehat{\Psi}_{4p+2}^{\,\alpha}(\hat{y}_1)\, \widehat{\Psi}_{4p+2}^{\,\beta}(\hat{y}_2) \,\rangle = \frac{a_p\, \delta^{\alpha\beta}}{|\hat{y}_{12}|^{(2p+1)\, d}}\, ,  
    \end{align}
where $a_p$ and $b_p$ are defined by;
\begin{align}\label{eq:a and b coeff}
    a_{p}\equiv 2^{6p+2}\, p!\, (N/2+1)_{p}\ ,\qquad  b_{p}\equiv 2^{6p+5}\, (p+1)!\, (N/2+1)_{p}\  .
\end{align}

\paragraph{Boundary three-point functions.} 
    \begin{align}
     \langle\,\widehat{\Psi}_{2}^\alpha(\hat{x})\, \widehat{\Psi}_{4p}(\hat{y}_1)\,\widehat{\Psi}_{4p+2}^{\,\beta}(\hat{y}_2)\,\rangle&=\frac{a_{p}\, \delta^{\alpha\beta}}{|\hat{x}-\hat{y}_{2}|^{d}|\hat{y}_{12}|^{2\, p\, d}}\ ,\label{eq:3pt ddd Dirichlet 1}\\
     \langle\,\widehat{\Psi}_{2}^\alpha(\hat{x})\, \widehat{\Psi}_{4p+2}^{\beta}(\hat{y}_1)\,\widehat{\Psi}_{4p+4}(\hat{y}_2)\,\rangle
    &=   \frac{b_{p}\, \delta^{\alpha\beta}}{|\hat{x}-\hat{y}_{2}|^{d}|\hat{y}_{12}|^{(2p+1)d}}\ ,\label{eq:3pt ddd Dirichlet 2}\\
      \langle\,\widehat{\Psi}_{6}^\alpha(\hat{x})\, \widehat{\Psi}_{4p+2}^{\,\beta}(\hat{y}_1)\,\widehat{\Psi}_{4p+4}(\hat{y}_2)\,\rangle&=\frac{4(N+6p+2)\, b_{p}\, \delta^{\alpha\beta}}{|\hat{x}-\hat{y}_{1}|^d |\hat{x}-\hat{y}_{2}|^{2d} |\hat{y}_{12}|^{2pd}}\ ,\label{eq:3pt ddd Dirichlet 3}\\
       \langle\,\widehat{\Psi}_{6}^\alpha(\hat{x})\, \widehat{\Psi}_{4p}(\hat{y}_1)\,\widehat{\Psi}_{ 4p+2}^{\,\beta}(\hat{y}_2)\,\rangle&=\frac{24p\, a_{p}\, \delta^{\alpha\beta}}{|\hat{x}-\hat{y}_{1}|^d |\hat{x}-\hat{y}_{2}|^{2d} |\hat{y}_{12}|^{(2p-1)d}} \label{eq:3pt ddd Dirichlet 4} \ .
\end{align}

\paragraph{Bulk-boundary-boundary three-point functions.}
\begin{align}
              \langle\,\Phi_{1}^\alpha(x)\, \widehat{\Psi}_{4p}(\hat{y}_1)\,\widehat{\Psi}_{4p+2}^{\,\beta}(\hat{y}_2)\,\rangle&= \frac{d-2}{2}\, 
              \frac{a_{p}\,  \delta^{\alpha\beta}\, x_{\perp}}{|x-\hat{y}_2|^d \,|\hat{y}_{12}|^{2pd}}\ ,\label{eq:3pt bdd Dirichlet 1}\\
                \langle\,\Phi_{1}^\alpha(x)\, \widehat{\Psi}_{4p+2}^{\beta}(\hat{y}_1)\,\widehat{\Psi}_{4p+4}(\hat{y}_2)\,\rangle
     &=  \frac{d-2}{2}\, \frac{b_{p}\, \delta^{\alpha\beta}\, x_{\perp}}{|x-\hat{y}_2|^d \,|\hat{y}_{12}|^{(2p+1)d}}\ ,\label{eq:3pt bdd Dirichlet 2}
     \end{align}
     \begin{align}
     \begin{aligned}
     &\langle\,\Phi_{3}^\alpha(x)\, \widehat{\Psi}_{4p}(\hat{y}_1)\,\widehat{\Psi}_{4p+2}^{\,\beta}(\hat{y}_2)\,\rangle& \\
     &\quad=-\frac{(N+2)\, (d-2)}{2^{d-1}}\frac{a_{p}\,\delta^{\alpha\beta}}{|x-\hat{y}_2|^d\,|\hat{y}_{12}|^{2pd}\,x_\perp^{d-3}}+\left(\frac{d-2}{2}\right)^3\frac{24p\,  a_{p}\, \delta^{\alpha\beta}\, x_{\perp}^3}{|x-\hat{y}_1|^d \,|x-\hat{y}_2|^{2d}\,|\hat{y}_{12}|^{(2p-1)d}}\ , &\label{eq:3pt bdd Dirichlet 3} 
     \end{aligned}
     \end{align}
     \begin{align}
     \begin{aligned}
     &\langle\,\Phi_{3}^\alpha(x)\, \widehat{\Psi}_{4p+2}^{\,\beta}(\hat{y}_1)\,\widehat{\Psi}_{4p+4}(\hat{y}_2)\,\rangle& \\
     &\quad=-\frac{(d-2)\, (N+2)}{2^{d-1}}\frac{b_{p}\,\delta^{\alpha\beta}}{|x-\hat{y}_2|^d\,|\hat{y}_{12}|^{(2p+1)d}\,x_\perp^{d-3}}+\left(\frac{d-2}{2}\right)^3\frac{4(N+6p+2)\,  b_{p}\, \delta^{\alpha\beta}\, x_{\perp}^3}{|x-\hat{y}_1|^d \,|x-\hat{y}_2|^{2d}\,|\hat{y}_{12}|^{2pd}}\, .&\label{eq:3pt bdd Dirichlet 4}
     \end{aligned}
\end{align}

\subsubsection{Boundary operator expansions in four dimensions}
We now elucidate the structure of the BOE of $\Phi_1^\alpha$ and $\Phi_3^\alpha$ in the four-dimensional free O$(N)$ model with the Dirichlet boundary condition. 
\paragraph{Boundary operator expansion of $\Phi_1^\alpha$.}
As in the Neumann case in section \ref{sec:Boundary operator expansion of Phi1}, only $\widehat{\Psi}_{2}^{\,\alpha}$ contribute to the BOE of $\Phi_1^\alpha$:
\begin{align}
    \Phi_1^\alpha(x) = x_\perp\,\widehat{\Psi}_{2}^\alpha(\hat{x})+(\text{descendants}) \label{eq:OPE of phi1 O(N) Dirichlet}\ .
\end{align}
This is consistent with the two-point functions \eqref{eq:2pt Phi Dir}, and also with the conformal block decomposition of the bulk-boundary-boundary three-point functions:\footnote{We implicitly used $G_2^{-2}(\upsilon)=\upsilon$ and \eqref{eq:a and b coeff}.}
\begin{align}
     \langle\,\Phi_1^\alpha(x)\, \widehat{\Psi}_{4p}(0)\,\widehat{\Psi}_{4p+2}^{\,\beta}(\infty)\,\rangle
        &=
         a_p\, \delta^{\alpha\beta}\, \frac{|x|^2}{x_\perp}\,G_2^{-2}\left(\frac{x_\perp^2}{|x|^2}\right)\, ,\label{eq:3pt bdd Dirichlet 1 exp}\\
      \langle\,\Phi_1^\alpha(x)\, \widehat{\Psi}_{4p+2}^{\beta}(0)\,\widehat{\Psi}_{4p+4}(\infty)\,\rangle
        &=
        b_p\, \delta^{\alpha\beta}\, \frac{|x|^2}{x_\perp}\, G_2^{-2}\left(\frac{x_\perp^2}{|x|^2}\right)\ .\label{eq:3pt bdd Dirichlet 2 exp}
\end{align}

\paragraph{Boundary operator expansion of $\Phi_3^\alpha$.}
It follows from the correlation functions (appendix \ref{app:Correlation functions in free theories in the Dirichlet case}) that the BOE of $\Phi_3^\alpha$ has the following operator contents:
\begin{align}\label{eq:btb OPE of phi3 all order; Dirichlet}
\begin{aligned}
  \Phi_3^{\alpha}(x) = \frac{1}{x_\perp}\,\widehat{\Psi}_2^\alpha(\hat{x})+
    \sum_{n=0}^{\infty}\,\frac{b(\Phi_3^\alpha,\widehat{\mathsf{Q}}_{2n+6}^\alpha)}{c(\widehat{\mathsf{Q}}_{2n+6}^\alpha,\widehat{\mathsf{Q}}_{2n+6}^\alpha)}\,x_\perp^{2n+3}\,\widehat{\mathsf{Q}}_{2n+6}^\alpha(\hat{x})+(\text{descendants}) \ .
\end{aligned}
\end{align}
Here $\widehat{\mathsf{Q}}_{6}^\alpha$ can be identified with $\widehat{\Psi}_6^\alpha$ and the ratio $b(\Phi_3^\alpha,\widehat{\mathsf{Q}}_{2n+6}^\alpha)/c(\widehat{\mathsf{Q}}_{2n+6}^\alpha,\widehat{\mathsf{Q}}_{2n+6}^\alpha)$ is subject to the following relations:\footnote{Note that, for $n\geq1$, $\widehat{\mathsf{Q}}_{2n+6}^\alpha$ is different from $\widehat{\Phi}_{2n+6}^\alpha$.}
\begin{align}\label{eq:bc_over_c_Dirichlet}
\begin{aligned}
    \frac{b(\Phi_3^\alpha,\widehat{\mathsf{Q}}_{2n+6}^\alpha)\,c(\widehat{\mathsf{Q}}_{2n+6}^\alpha,\widehat{\Psi}_{4p},\widehat{\Psi}_{4p+2}^{\,\beta})}{c(\widehat{\mathsf{Q}}_{2n+6}^\alpha,\widehat{\mathsf{Q}}_{2n+6}^{\alpha})}&=24p\, a_p\, \frac{(-1)^n\,(2)_n (4)_n}{(n+9/2)_n \, n!}\, \delta^{\alpha\beta}\ ,\\
    \frac{b(\Phi_3^\alpha,\widehat{\mathsf{Q}}_{2n+6}^\alpha)\,c(\widehat{\mathsf{Q}}_{2n+6}^\alpha,\widehat{\Psi}_{4p+2}^{\,\beta},\widehat{\Psi}_{4p+4})}{c(\widehat{\mathsf{Q}}_{2n+6}^\alpha,\widehat{\mathsf{Q}}_{2n+6}^{\alpha})}&=4(N+6p+2)\, b_p\, \frac{(-1)^n\,(2)_n (4)_n}{(n+9/2)_n \, n!}\, \delta^{\alpha\beta}\ .
\end{aligned}
\end{align}
We can convince ourselves of the validity of \eqref{eq:btb OPE of phi3 all order; Dirichlet} and \eqref{eq:bc_over_c_Dirichlet} by the following arguments.

By Wick's theorem or looking at the two-point functions \eqref{eq:2pt Phi Dir} and \eqref{eq:Dir btb 2pt Phi3}, we infer the following BOE of $\Phi_3^\alpha$:
\begin{align}
    \Phi_3^\alpha(x)\supset -\frac{N+2}{4\,x_\perp}\,\widehat{\Psi}_{2}^\alpha(\hat{x}) + x_\perp^3\,\widehat{\Phi}_{6}^\alpha(\hat{x}) \label{eq:OPE of phi3 O(N) Dirichlet}\ .
\end{align}
To proceed, we focus on the bulk-boundary-boundary three-point functions involving $\Phi_3^\alpha$ \eqref{eq:3pt bdd Dirichlet 3} and \eqref{eq:3pt bdd Dirichlet 4}:
\begin{align}
     \langle\,\Phi_3^\alpha(x)\, \widehat{\Psi}_{4p}(0)\,\widehat{\Psi}_{4p+2}^{\,\beta}(\infty)\,\rangle&=\frac{|x|^2}{x_\perp^3}\left[-\frac{N+2}{4}a_p\, \delta^{\alpha\beta}\frac{x_\perp^2}{|x|^2}+24p\, a_p\, \frac{x_\perp^6}{|x|^6}\right]\ ,\label{eq:3pt bdd Dirichlet 3 exp} \\
       \langle\,\Phi_3^\alpha(x)\, \widehat{\Psi}_{4p+2}^{\,\beta}(0)\,\widehat{\Psi}_{4p+4}(\infty)\,\rangle&=\frac{|x|^2}{x_\perp^3}\left[-\frac{N+2}{4}b_p\, \delta^{\alpha\beta}\frac{x_\perp^2}{|x|^2}+4(N+6p+2)\, b_p\, \delta^{\alpha\beta}\, \frac{x_\perp^6}{|x|^6}\right]\ , \label{eq:3pt bdd Dirichlet 4 exp}
\end{align}
whose conformal block expansions are
\begin{align}\label{eq:3pt bdd Dirichlet 3 exp ast}
\begin{aligned}
     \langle\,\Phi_3^\alpha(x)\, \widehat{\Psi}_{4p}(0)&\,\widehat{\Psi}_{4p+2}^{\,\beta}(\infty)\,\rangle
     =
            -\frac{N+2}{4}\,a_p\, \delta^{\alpha\beta}\,\frac{|x|^2}{x_\perp^3}\, G^{-2}_{2}\left(\frac{x_\perp^2}{|x|^2}\right)\\
        &\qquad 
            +           \frac{|x|^2}{x_\perp^3}\, \sum_{n=0}^{\infty}\,24\,p\,a_p\,\delta^{\alpha\beta}\,\frac{(-1)^n\,(2)_n (4)_n}{(n+9/2)_n\, n!}\, G^{-2}_{2n+6}\left(\frac{x_\perp^2}{|x|^2}\right)\ ,
\end{aligned}
\end{align}
and 
\begin{align}\label{eq:3pt bdd Dirichlet 4 exp ast}
\begin{aligned}
     \langle\,\Phi_3^\alpha(x)\, \widehat{\Psi}_{4p+2}^{\,\beta}&(0)\,\widehat{\Psi}_{4p+4}(\infty)\,\rangle=
            -\frac{N+2}{4}\,b_p\, \delta^{\alpha\beta}\,\frac{|x|^2}{x_\perp^3}\, G^{-2}_{2}\left(\frac{x_\perp^2}{|x|^2}\right)\\
        &\qquad 
            +           \frac{|x|^2}{x_\perp^3}\, \sum_{n=0}^{\infty}\,4(N+6p+2)\, b_p\,\,\delta^{\alpha\beta}\,\frac{(-1)^n\,(2)_n (4)_n}{(n+9/2)_n\, n!}\, G^{-2}_{2n+6}\left(\frac{x_\perp^2}{|x|^2}\right)\ .
\end{aligned}
\end{align}
Comparing \eqref{eq:3pt bdd Dirichlet 3 exp ast} and \eqref{eq:3pt bdd Dirichlet 4 exp ast} with \eqref{eq:conformal block expansion main}, we obtain \eqref{eq:bc_over_c_Dirichlet}.

In what follows, we calculate the anomalous dimensions of the composite operators $\widehat{\CW}_{4p}$ and $\widehat{\CW}_{4p+2}^{\,\alpha}$ which reduce to $\widehat{\Psi}_{4p}$ and $\widehat{\Psi}_{4p+2}^{\, \alpha}$ in the free limit, via both the diagrammatic and axiomatic methods.   

\subsection{Diagrammatic approach}\label{app: Diagrammatic Calc Dirichlet}
We define the renormalization factors $Z_{4p}$ and $Z_{4p+2}$ as follows:
\begin{align}\label{eq:wave function ren Dirichlet}
                \begin{aligned}
                    \widehat{\CW}_{4p}=Z_{4p}^{-1}\,\widehat{\Psi}_{4p} \ ,\quad \widehat{\CW}_{4p+2}^{\,\alpha}=Z_{4p+2}^{-1}\, \widehat{\Psi}_{4p+2}^{\,\alpha}\  .
                \end{aligned}
            \end{align}
The conformal dimension of $\widehat{\CW}_{4p}$ and $\widehat{\CW}_{4p+2}^{\,\alpha}$ are denoted by $\widehat{\Delta}_{4p}$ and $\widehat{\Delta}_{4p+2}$, respectively.
We focus on the leading anomalous dimension $\widehat{\gamma}_{n,1}$ defined through:
\begin{align}\label{eq:anomalous dimensions Dirichlet def}
     \widehat{\Delta}_{n}=\frac{n\, d}{4}+\widehat{\gamma}_{n}\ ,\qquad \widehat{\gamma}_n=\widehat{\gamma}_{n,1}\,\epsilon+\widehat{\gamma}_{n,2}\,\epsilon^2+\cdots\ ,\qquad \widehat{\gamma}_{n}\equiv\beta_{\lambda} \left. \frac{\text{d}\,  \ln Z_{n}}{\text{d} \lambda}\right|_{\lambda=\lambda_\ast} \ .
 \end{align}
By evaluating the following correlation functions at one-loop level
\begin{align}\label{eq: two-point correlation function, interacting, Dirichlet}
    \begin{aligned}
         \langle\, \widehat{\Psi}_{4p}(\hat{y})\,\widehat{\Psi}_{4p}(0) \,\rangle\, \ ,\qquad  \langle\, \widehat{\Psi}_{4p+2}^{\,\alpha}(\hat{y})\,\widehat{\Psi}_{4p+2}^{\,\beta}(0) \,\rangle \ ,
    \end{aligned}
\end{align}
we obtain two recursion relations similar to \eqref{eq: recursion relation between Z_2p and Z_{2p-1}} and \eqref{eq: recursion relation between Z_{2p-1} and Z_{2p-2}}:
\begin{align} \label{eq: recursion relation Dirichlet}
\begin{aligned}
    \delta Z_{4p}-\delta Z_{4p-2}=-\frac{N+12p-10}{6\,\epsilon}\, \pi^2\, \lambda + O(\lambda^2, \epsilon^0) \ , \\
     \delta Z_{4p-2}-\delta Z_{4p-4}=-\frac{12p-N-14}{6\,\epsilon}\, \pi^2\, \lambda +O(\lambda^2, \epsilon^0)\ .
\end{aligned}
\end{align}
Finally, by solving them under $\delta Z_{0}=0$, we get
\begin{align}\label{eq: Diagrammatic result, Dirichlet, App}
     \widehat{\gamma}_{4p,1}
        =
            \frac{6p\,(p-1)}{N+8}\ ,\qquad 
    \widehat{\gamma}_{4p+2,1}
        =
            \frac{12\, p^2 -N-2}{2(N+8)}\ .
\end{align}

\subsection{Axiomatic approach}
Next, we use the axiomatic framework to derive the leading anomalous dimensions of the boundary local operators.

\subsubsection{Lowest-lying boundary local operator}
We first focus on the lowest-lying boundary local operator $\widehat{\CW}_{2}^{\,\alpha}$ that approaches $\widehat{\Psi}_2^{\,\alpha}$ as $\epsilon\to0$.
The derivation is completely parallel to the Neumann case. We start with the boundary OPE of $W_1^\alpha$:
\begin{align}
    W_1^\alpha(x)
        \supset D\,\frac{1}{|x_\perp|^{\Delta_1-\widehat{\Delta}_2}}\,\widehat{\CW}_{2}^{\,\alpha}(\hat{x}) \ ,\qquad D=1+O(\epsilon) \ .
\end{align}
For this to match with \eqref{eq:OPE of phi1 O(N) Dirichlet}, we have;
\begin{align}\label{eq:Dirichlet lowest matching}
    D=1+O(\epsilon)\ ,\qquad \widehat{\Delta}_2=2+O(\epsilon)\ .
\end{align}
The equation of motion \eqref{eq:classical EoM O(N)} and the BOE lead to
\begin{align}
\begin{aligned}
     W_3^\alpha(x)
        &=
        \frac{1}{\kappa}\,\Box\,W_1^\alpha(x)\\
        &\supset \frac{D}{\kappa}\,\frac{(\Delta_1-\widehat{\Delta}_2)(\Delta_1-\widehat{\Delta}_2+1)}{|x_\perp|^{\Delta_1-\widehat{\Delta}_2+2}}\,\widehat{\CW}_{2}^{\,\alpha}(\hat{x}) \ .  
\end{aligned}
\end{align}
Since this should match with \eqref{eq:OPE of phi3 O(N) Dirichlet} in $\epsilon\to0$, we have the following equation:
\begin{align}
- \frac{N+2}{4}\,\kappa=(\widehat{\Delta}_2-\Delta_1)(\widehat{\Delta}_2-\Delta_1-1) +O(\epsilon^2) \ .
\end{align}
Hence, the solution compatible with \eqref{eq:Dirichlet lowest matching} is given by 
\begin{align}\label{eq:anomalous dim lowest Dir}
    \begin{aligned}
    \widehat{\Delta}_2
        &=
        \frac{d}{2}-\frac{N+2}{2\,(N+8)}\,\epsilon+O(\epsilon^2) \\
        &=
        2 -\frac{N+5}{N+8}\,\epsilon+O(\epsilon^2)\ ,
    \end{aligned}
\end{align}
which reproduces the known diagrammatic result \cite[equation (3.8)]{McAvity:1995zd}.

\subsubsection{Boundary composite operators}\label{app:Dir higher order}
We proceed in a similar manner to section \ref{sec:Neu higher order} to calculate the conformal dimensions of the boundary composite operators $\widehat{\CW}_{4p}$ and $\widehat{\CW}_{4p+2}^{\,\alpha}$ that reduce to $\widehat{\Psi}_{4p}$ and $\widehat{\Psi}_{4p+2}^{\,\alpha}$ in $\epsilon\to0$.

\paragraph{Boundary operator expansion of $W_1^\alpha$.}
The boundary local operators appearing in the BOE of $W_{1}^{\alpha}$ with the Dirichlet boundary condition are different from those with the Neumann boundary case \eqref{eq:OPE of W1 all order}:
\begin{align}\label{eq:OPE of W1 all order;Dirichlet}
    \begin{aligned}
        W_1^\alpha(x)
            &=
                \frac{D}{|x_\perp|^{\Delta_1-\widehat{\Delta}_2}}\,\widehat{\CW}_2^{\,\alpha}(\hat{x}) 
          \\
            &\qquad    +
  \sum_{n=0}^{\infty}\,\frac{b(W_1^\alpha,\widehat{\mathsf{Q}}_{2n+6}^{\prime\,\alpha})/c(\widehat{\mathsf{Q}}_{2n+6}^{\prime\,\alpha},\widehat{\mathsf{Q}}_{2n+6}^{\prime\,\alpha})}{|x_\perp|^{\Delta_1-\widehat{\Delta}'_{2n+6}}}\,\widehat{\mathsf{Q}}_{2n+6}^{\prime\,\alpha}(\hat{x})+(\text{descendants})\ ,
    \end{aligned}
\end{align}
where 
\begin{align}
    \begin{aligned}
        \lim_{\epsilon\rightarrow 0}\,\widehat{\mathsf{Q}}_{2n+6}^{\prime\,\alpha}(\hat{x})=\widehat{\mathsf{Q}}_{2n+6}^\alpha(\hat{x}) \ , \qquad  \lim_{\epsilon \rightarrow 0}\,\widehat{\Delta}'_{2n+6}=2n+6\ ,
    \end{aligned}
\end{align}
and $\widehat{\mathsf{Q}}_{6}^{\prime\,\alpha}$ can be identified with $\widehat{\CW}_6^{\,\alpha}$.
In a similar manner to the Neumann case, the equation of motion \eqref{eq:classical EoM O(N)} fixes the $O(\epsilon)$ coefficient of $b(W_1^\alpha,\widehat{\mathsf{Q}}_{2n+6}^{\prime\,\alpha})/c(\widehat{\mathsf{Q}}_{2n+6}^{\prime\,\alpha},\widehat{\mathsf{O}}_{2n+3}^{\prime\,\alpha})$:
\begin{align}\label{eq:W1 to higher order OPE coeff;Dirichlet}
\frac{b(W_1^\alpha,\widehat{\mathsf{Q}}_{2n+6}^{\prime\,\alpha})}{c(\widehat{\mathsf{Q}}_{2n+6}^{\prime\,\alpha},\widehat{\mathsf{Q}}_{2n+6}^{\prime\,\alpha})}=\frac{\kappa}{(2n+4)(2n+5)}\frac{b(\Phi_3^\alpha,\widehat{\mathsf{Q}}_{2n+6}^\alpha)}{c(\widehat{\mathsf{Q}}_{2n+6}^\alpha,\widehat{\mathsf{Q}}_{2n+6}^\alpha)}+O(\epsilon^2)\ .
\end{align}

\paragraph{Bulk-boundary-boundary three-point functions involving $W_1^\alpha$.}\label{sec:bbb 3pt W1 Dirichlet}
By exploiting the BOE of $W_1^\alpha$ derived in the last paragraph, one can evaluate the bulk-boundary-boundary three-point functions.
First, consider $\langle\, W_1^\alpha(x)\, \widehat{\CW}_{4p}(0)\,\widehat{\CW}_{4p+2}^{\,\beta}(\infty)\,\rangle$:
\begin{align}\label{eq: 3pt bdd, WilsonFisher;Dirichlet}
    \begin{aligned}
    \langle\,& W_1^\alpha(x)\, \widehat{\CW}_{4p}(0)\,\widehat{\CW}_{4p+2}^{\,\beta}(\infty)\,\rangle=\frac{1}{|x_\perp|^{\Delta_1}\, |x|^{\widehat{\Delta}_{4p}-\widehat{\Delta}_{4p+2}}}\\
    &\qquad\cdot\left[ D\cdot c(\widehat{\CW}_2^{\,\alpha},\widehat{\CW}_{4p},\widehat{\CW}_{4p+2}^{\,\beta})\,G^{\widehat{\Delta}_{4p}-\widehat{\Delta}_{4p+2}}_{\widehat{\Delta}_2}(\upsilon)\right.\\
    &\qquad\qquad\left.  +\sum_{n=0}^{\infty}\,\frac{b(W_1^\alpha,\widehat{\mathsf{Q}}_{2n+6}^{\prime\,\alpha})\,c(\widehat{\mathsf{Q}}_{2n+6}^{\prime\,\alpha},\widehat{\CW}_{4p},\widehat{\CW}_{4p+2}^{\, \beta})}{c(\widehat{\mathsf{Q}}_{2n+6}^{\prime\,\alpha},\widehat{\mathsf{Q}}_{2n+6}^{\prime\,\alpha})}\,G^{\widehat{\Delta}_{4p}-\widehat{\Delta}_{4p+2}}_{\widehat{\Delta}'_{2n+6}}(\upsilon)\right]\ .
    \end{aligned}
\end{align}
where $\upsilon$ is defined by $\upsilon\equiv |x_{\perp}|^2 /|x|^2$. The first term in the RHS of \eqref{eq: 3pt bdd, WilsonFisher;Dirichlet} can be evaluated as
\begin{align}
\begin{aligned}
    G^{\widehat{\Delta}_{4p}-\widehat{\Delta}_{4p+2}}_{\widehat{\Delta}_2}(\upsilon)&=\upsilon^{\widehat{\Delta}_2/2}\,{}_2F_1\left(\frac{\widehat{\gamma}_{2,1}+\widehat{\gamma}_{4p,1}-\widehat{\gamma}_{4p+2,1}}{2}\,\epsilon\, ,2; 3/2;\upsilon\right)+O(\epsilon^2) \\
    &= \upsilon^{\widehat{\Delta}_2/2} + \frac{\widehat{\gamma}_{2,1}+\widehat{\gamma}_{4p,1}-\widehat{\gamma}_{4p+2,1}}{3} \,\epsilon\, \upsilon^2\,h(\upsilon)\, +O(\epsilon^2)\ ,
\end{aligned}
\end{align}
where we introduced
\begin{align}\label{eq:function h def}
    h(\upsilon)={}_2F_1(1,2;5/2;\upsilon)+{}_2F_1(1,1;5/2;\upsilon)=\sum_{n=0}^{\infty}\,\frac{(1)_n\,(n+2)}{(5/2)_n}\,\upsilon^n\ .
\end{align}
Using \eqref{eq:bc_over_c_Dirichlet} and \eqref{eq:W1 to higher order OPE coeff;Dirichlet}, the second term in RHS of \eqref{eq: 3pt bdd, WilsonFisher;Dirichlet} becomes\footnote{In particular, we used the relation
\begin{align*}
 \sum_{n=0}^{\infty}\,\frac{(-1)^n\, (n+1)\,(4)_n}{(n+2)(2n+5)(n+9/2)_n} \, \upsilon^{n+1}\,{}_2F_1(n+2,n+4;2n+11/2;\upsilon)=\frac{1}{12}\,\left[ -2 + \sum_{n=0}^\infty\, \frac{(1)_{n}\,(n+2)}{(5/2)_{n}}\,\upsilon^{n}\right]\ .
\end{align*}
}
\begin{align} \sum_{n=0}^{\infty}\,\frac{b(W_1^\alpha,\widehat{\mathsf{Q}}_{2n+6}^{\prime\,\alpha})\,c(\widehat{\mathsf{Q}}_{2n+6}^{\prime\,\alpha},\widehat{\CW}_{4p},\widehat{\CW}_{4p+2}^{\, \beta})}{c(\widehat{\mathsf{Q}}_{2n+6}^{\prime\,\alpha},\widehat{\mathsf{Q}}_{2n+6}^{\prime\,\alpha})}\,G^{\widehat{\Delta}_{4p}-\widehat{\Delta}_{4p+2}}_{\widehat{\Delta}'_{2n+6}}(\upsilon)
    =
    \kappa \, p \, a_p \, \delta^{\alpha\beta} \,  \upsilon^2\,\left[ -2 + h(\upsilon)\right]+O(\epsilon^2)\ .
    \end{align}
Plugging these two into \eqref{eq: 3pt bdd, WilsonFisher;Dirichlet}, we find
\begin{align}\label{eq:W1 3pt expansion 1; Dirichlet}
    \begin{aligned}
    \langle\,& W_1^\alpha(x)\, \widehat{\CW}_{4p}(0)\,\widehat{\CW}_{4p+2}^{\,\beta}(\infty)\,\rangle\\
    &=c(\widehat{\CW}_2^{\,\alpha},\widehat{\CW}_{4p},\widehat{\CW}_{4p+2}^{\,\beta})\, \frac{D}{x_\perp^{\Delta_1 -\widehat{\Delta}_2}|x|^{\widehat{\Delta}_2+\widehat{\Delta}_{4p}-\widehat{\Delta}_{4p+2}}} \\
     &\quad +\frac{a_p\, \delta^{\alpha\beta}\,|x_\perp|^3}{3\,|x|^2}\, \left\{
 -6\kappa\, p+\left[(\widehat{\gamma}_{2,1}+\widehat{\gamma}_{4p,1}-\widehat{\gamma}_{4p+2,1})\,\epsilon\,+3\,\kappa\, p\right]\cdot h(\upsilon) \right\}
+O(\epsilon^2)\ .
    \end{aligned}
\end{align}
Similarly for $\langle\, W_1^\alpha\, \widehat{\CW}^{\, \beta}_{4p+2}\,\widehat{\CW}_{4p+4}\,\rangle$, we end up with
{\small\begin{align}\label{eq:W1 3pt expansion 2; Dirichlet}
    \begin{aligned}
    \langle\,& W_1^\alpha(x)\, \widehat{\CW}^{\, \beta}_{4p+2}(0)\,\widehat{\CW}_{4p+4}(\infty)\,\rangle\\
    &=c(\widehat{\CW}_2^{\,\alpha},\widehat{\CW}^{\, \beta}_{4p+2},\widehat{\CW}_{4p+4})\,\frac{D}{x_\perp^{\Delta_1 -\widehat{\Delta}_2}|x|^{\widehat{\Delta}_2+\widehat{\Delta}_{4p+2}-\widehat{\Delta}_{4p+4}}} \\
     &\quad +\frac{b_p\, \delta^{\alpha\beta}\,|x_\perp|^3}{3\,|x|^2}\,\left\{
 -(N+6p+2)\,\kappa+\left[(\widehat{\gamma}_{2,1}+\widehat{\gamma}_{4p+2,1}-\widehat{\gamma}_{4p+4,1})\,\epsilon\,+\frac{N+6p+2}{2}\,\kappa\, \right]\cdot h(\upsilon)\right\} +O(\epsilon^2)\ .
    \end{aligned}
\end{align}}

\paragraph{Constraint from analyticity.}\label{sec:Requiring holomorphicity:Dirichlet}
We notice that \eqref{eq:W1 3pt expansion 1; Dirichlet} and \eqref{eq:W1 3pt expansion 2; Dirichlet} have the unphysical singularity coming from \eqref{eq:function h def}:
\begin{align}
    h(\upsilon) \xrightarrow[|\hat{x}|\sim 0]{} \frac{3\pi}{4}\cdot\frac{|x_{\perp}|}{|\hat{x}|}\, +(\text{less singular terms})\ .
\end{align}
The absence of the singularity leads us to the following relations between anomalous dimensions:
\begin{align}\label{eq:Dirichlet req from regularity}
    (\widehat{\gamma}_{2,1}+\widehat{\gamma}_{4p,1}-\widehat{\gamma}_{4p+2,1})\,\epsilon+ 3\,\kappa\,p 
        &=
            0+O(\epsilon^2)\ ,\\
    ( \widehat{\gamma}_{2,1}+\widehat{\gamma}_{4p+2,1}-\widehat{\gamma}_{4p+4,1})\,\epsilon+ \kappa\,\frac{N+6p+2}{2}
        &=
            0+O(\epsilon^2)\ .
\end{align}
By solving these recursion relations with the initial condition $\widehat{\gamma}_{2,1}=-(N+2)/2(N+8)$ \eqref{eq:anomalous dim lowest Dir}, we reproduce the same results as those obtained by the diagrammatic approach \eqref{eq: Diagrammatic result, Dirichlet, App}.

\bibliographystyle{JHEP}
\bibliography{DCFT}

\end{document}